\documentclass[aps,amsfonts,amsmath,prd,preprint,nofootinbib]{revtex4}
\newcommand{\beq}{\begin{equation}}
\newcommand{\eeq}{\end{equation}}

\usepackage{epsfig,bbm,cancel,ulem}
\usepackage[breaklinks=true]{hyperref}
\usepackage{xcolor}
\usepackage{wasysym}
\usepackage{graphicx}
\usepackage{epstopdf}
\usepackage{color}
\usepackage{bm}
\newcommand {\bea}{\begin{eqnarray}}
\newcommand {\eea}{\end{eqnarray}}
\newcommand {\be}{\begin{equation}}
\newcommand {\ee}{\end{equation}}

\begin{document}
	\def\Journal#1#2#3#4{{ #1} {\bf #2}, #3 (#4) }
	\def\RPP{{Rep. Prog. Phys}}
	\def\PRC{{Phys. Rev. C}}
	\def\PRD{{Phys. Rev. D}}
	\def\ZPA{{Z. Phys. A}}
	\def\NPA{{Nucl. Phys. A}} 
	\def\JPG{{J. Phys. G }}
	\def\PRL{{Phys. Rev. Lett}}
	\def\PR{{Phys. Rep.}}
	\def\PREV{{Phys. Rev.}}
	\def\PRX{{Phys. Rev. X}}
	\def\PLB{{Phys. Lett. B}}
	\def\AP{{Ann. Phys (N.Y.)}}
	\def\EPJA{{Eur. Phys. J. A}}
	\def\NP{{Nucl. Phys}}  
	\def\RMP{{Rev. Mod. Phys}}
	\def\IJMPE{{Int. J. Mod. Phys. E}}
	\def\AJ{{Astrophys. J}}
	\def\AJL{{Astrophys. J. Lett}}
	\def\AA{{Astron. Astrophys}}
	\def\ARAA{{Annu. Rev. Astron. Astrophys}}
	\def\MPLA{{Mod. Phys. Lett. A}}
	\def\ARNPS{{Annu. Rev. Nuc. Part. Sci}}
	\def\LRR{{Living. Rev. Relativity}}
	\def\CARAA{{Class. Ann. Rev. Astron. Astrophys.}}
	\def\EPJC{{Eur. Phys. J. C}}
	\def\cqg{{Class. Quantum Grav.}}
	\def\mon{{Mon. Not. R. Astron. Soc.}}
        \def\AJSS{{Astrophys. J. Suppl. Ser.}}	
\title{$2.6 M_\odot$ Compact Object and Neutron Stars within Eddington-Inspired Born-Infeld Theory of Gravity}
	
\author{I. Prasetyo}
\email{ilham.prasetyo@sci.ui.ac.id}

\author{H. Maulana}
\email{haris.maulana@alumni.ui.ac.id}

\author{H. S. Ramadhan}
\email{hramad@sci.ui.ac.id}

\author{A. Sulaksono}
\email{anto.sulaksono@sci.ui.ac.id}

\affiliation{Departemen Fisika, FMIPA, Universitas Indonesia, Depok 16424, Indonesia. }
	
\def\changenote#1{\footnote{\bf #1}}
	
\begin{abstract}
In the context of whether a massive compact object recently observed in the GW190814 event is a neutron star (NS) or not, we have studied the role of the parameters $\kappa$ and $\Lambda_c$ of the Eddington-inspired Born-Infeld (EiBI) gravity theory on the NS mass-radius relation, moment of inertia, and tidal deformability. 
The results are compared to recent observation constraints extracted from the analysis of NS observation data.
The NS core equation of state (EoS) is calculated using the relativistic mean-field model with the G3 parameter set. In the hyperon sector, the SU(3) and hyperon potential depths are used to determine the hyperon coupling constants. For the inner and outer crusts, we use the crust EoS from Miyatsu {\it et al.} (2013). We also maintain the sound speed to not exceed $c$/$\sqrt{3}$ at high densities. 
We have found that, in general, the NS mass significantly depends on the value of $\kappa$, and the radius $R$ is sensitive to the value of $\Lambda_c$. Moreover, as $\Lambda_c$ is equal to zero or less than the accepted bound of the cosmological constant, the NS within the EiBI theory is compatible with observation constraints, including $2.0 M_\odot$ mass, canonical radius $R_{1.4 M_{\odot}}$, moment of inertia, and tidal deformation. Our investigation also reveals that the $2.6 M_\odot$ mass compact object and current observational constraint of canonical radius $R_{1.4 M_{\odot}}$ can simultaneously be satisfied only when the $\Lambda_c$ value is unphysically too large and negative. Therefore, \textcolor{black}{ within the spesific EoS employed in this work,} we conclude that the secondary object with $2.6 M_\odot$ observed in the GW190814 event~\cite{Abbott:2020khf} is not likely a static (or a slow-rotating) NS within the EiBI gravity theory.
\end{abstract}

	\maketitle
	\thispagestyle{empty}
	\setcounter{page}{1}

\section{INTRODUCTION}
\label{intro}
Recently, the most known problem in compact objects is the nature of $2.50-2.67 M_\odot$ massive secondary objects detected in the gravitational wave (GW) by the LIGO and Virgo collaboration in their GW190814 event~\cite{Abbott:2020khf}. These objects have no measurable signature of tidal deformation, and there is no electromagnetic counterpart on the gravitational wavefront. Considerable discussions have been published about this object, such as whether it is a light black hole (BH)~\cite{Abbott:2020khf,Fattoyev:2020cws,Nathanail:2021tay,Li:2020ias}, a fast rotating neutron star (NS)~\cite{Abbott:2020khf,Zhang:2020zsc,Most:2020bba,Zhou:2020xan,Biswas:2020xna}, a quark star~\cite{ Yang:2021sqg,Rather:2021yxo,Roupas:2020nua,Sedrakian:2020kbi,Cao:2020zxi,Bombaci:2020vgw,Zhang:2020jmb}, or a hybrid star~\cite{Zhang:2020dfi,Rather:2020lsg}. However, one could not exclude the possibility that the secondary object of GW190814 can be a super-massive static or at least a slow rotating NS~\cite{Fattoyev:2020cws,Huang:2020cab,Das:2020dcq}. The latter possibility has triggered discussions on the appropriate type of equation of state (EoS) of the super-massive NS that satisfies observational constraints~\cite{Drischler:2020fvz,Dexheimer:2020rlp,Kanakis-Pegios:2020kzp,Tan:2020ics,Huang:2020cab,Das:2020dcq}. Furthermore, studies have discussed the anisotropic pressure to calculate the upper mass limit~\cite{Horvath:2020lwj,Demircik:2020jkc,Roupas:2020jyv}, the possibility of studying primordial BH~\cite{Lehmann:2020bby}, the indication of a dark matter candidate called a mirror world~\cite{Beradze:2021fdw}, and the use of modified gravity to explain GW events~\cite{Barros:2021jbt,Moffat:2020jic,Nunes:2020cuz,Astashenok:2020qds}.

Here, we note some progress related to the observations of NS properties. The accurate measurements of massive pulsars, such as PSR J0348+0432, PSR J0740+6620, and J6114-2230\cite{Demorest:2010bx,Fonseca:2016tux,Arzoumanian:2017puf,Cromartie:2019kug,Antoniadis2013}, provide a maximum NS mass limit of approximately 2.0 $M_\odot$. The X-ray measurements of emission from the hot spots on the NS surface with the Neutron star Interior Composition Explorer (NICER)~\cite{Watts2016} can simultaneously offer information on the mass and radius of the selected pulsars. Recently, NICER reported mass and radius constraints for its first target PSR, i.e., PSR J0030+0451~\cite{Guillot2019,Bogdanov2019a,Bogdanov2019b}. GW observations of NS coalescence by the LIGO and Virgo collaborations can measure the tidal deformability of NSs. This novel probe can investigate a wide range of NS mass and the corresponding central density~\cite{Abbott2017,Abbott2018,Abbott2019,Abbott:2020khf}. Two GW signals from the coalescence of binary NSs have been recently reported, i.e., GW170817~\cite{Abbott2017,Abbott2018}, and GW190425~\cite{Abbott:2020khf}. These results provide a stringent constraint to the NS EoS and canonical NS mass radius.
Furthermore, some studies have been performed by systematically examining these NS observable measurements and other observable measurements, such as NS \textcolor{black}{moment of inertia} and nuclear properties, to extract the accurate information of the properties of NS EoS~\cite{Landry:2020vaw,Jiang:2019rcw,Kumar:2019xgp,Lim:2018xne,Lattimer:2004nj,Breu:2016ufb}. Furthermore, one uses non-relativistic or relativistic models to describe NS matter. Several NS matter models have been proposed, including the relativistic mean-field (RMF) models. Dutra {\it et al.}~\cite{Dutra2014} reported only 34 from 263 RMF parameter sets that satisfy nuclear matter constraints. Furthermore, in isotropic NSs without hyperons, only 15 among 35 parameter sets predicted the NS maximum mass of approximately 2.0 $M_\odot$. However, if hyperons and other exotic particles are included, then none of them satisfy the later constraint\footnote{See Ref.~\cite{Odilon2019} and the references therein for details.}. The latter is known in the literature as ``the hyperon puzzle.'' To this end, we need to underline that the apparent tension between nuclear physics, presented by EoS, and some observation results of NS EoS models should be relatively stiff to produce an NS maximum mass of approximately 2.0 $M_\odot$. Moreover, recent NS canonical radii, such as those predicted by GW170817, have soft EoSs. In addition, a recent study has shown that by introducing anisotropic pressure in NS, this issue, related to simultaneously fulfilling high maximum mass and short canonical NS mass-radius constraints~\cite{Biswas:2020puz}, could be resolved~\cite{Rahmansyah:2020gar}. However, if the secondary object of GW190814 is indeed a static or slow rotating NS, then the tension may still tight.

The Eddington-inspired Born-Infeld (EiBI) theory has attracted considerable attention due to its distinctive features as compared to those of general relativity (GR)~\cite{Vollick:2003qp,Banados:2010ix,Pani:2011mg,Pani:2012qb,Delsate12,Harko:2013wka,Wibisono:2017dkt,Danarianto:2019mxf,Rosyadi:2019hdb}. The EiBI theory, proposed for the first time by Banados and Ferreira~\cite{Banados:2010ix}, is a fusion of the Palatini approach and a gravitational analog of a nonlinear theory of electrodynamics known as the Born-Infeld theory\footnote{The reviews of the corresponding theory and applications of the EiBI gravity theory can be found in Refs.~\cite{JHOR2017,Berti_etal2015} and the references therein.}. In the astrophysical context, the EiBI theory is interesting because it opens up the possibility to increase the maximum mass $M$ of a nonrotating compact object, such as NSs, by increasing the parameter $\kappa$~\cite{Qauli:2016vza,Prasetyo:2017hrb,Qauli:2017ntr}.
Another parameter $\lambda$ in EiBI, corresponding to the cosmological constant $\Lambda_c$ by $\lambda=\kappa\Lambda_c+1$ relation, is usually set to unity for most cases for compact object studies, including NSs. However, the problem with $\lambda=1$ is that when the mass increases, the radius $R$ also increases. 
For the $\lambda=1$ case, stars' moment of inertia was discussed in Ref.~\cite{Pani:2012qb}. Furthermore, in Ref.~\cite{Pani:2012qb} it was discussed that a regular solution for compact stars with $\kappa >$ 0 always exists, and the corresponding stars have a maximum compactness of $\frac{GM}{R} \sim$ 0.3, which is roughly independent from $\kappa$. There is also a requirement called the collapse constraint, i.e., the compact stars exist if the requirement $\kappa \Delta < 0$ is satisfied, with $\Delta$ of
	\begin{equation}
	\Delta=(P_c \kappa - 3 \kappa \rho_c -4)(1 + \kappa \rho_c)-\kappa (1-\kappa P_c)(P_c+\rho_c) \frac{ d \rho (P_c)}{d P_c},
	\nonumber
	\end{equation}
where $P_c$ and $\rho_c$ are the central pressure and density of the stars, respectively. Hence, if the EoS is thermodynamically consistent, then the onset of the star's stability region in the EiBI theory depends only on $P_c$ and $\kappa$. As regards to the stellar stability of the stars within the EiBI theory, Sham {\it et al.} (2012)~\cite{SLL2012} showed that the standard results of stellar stability still hold in the EiBI theory, where for a sequence of stars with the same EoS, the fundamental mode $\omega^2$ passes through zero at a central density corresponding to the maximum-mass configuration, which is similar to that found in GR. Therefore, the corresponding point marks the boundary of the onset of instability, where the stellar models with central densities less than the corresponding critical points are stable. The EiBI theory also shows a singularity associated with the phase transition matter for a negative $\kappa$ due to the appearance of discontinuity in the energy density around the transition region~\cite{Sham2013}. The curvature singularities appearing at the surface of compact stars within the EiBI theory for polytropic EoSs have already been discussed in Refs.~\cite{BSM2008,PS2012,PSV2013,Kim2014}. \textcolor{black}{There is a discussion related to the tidal deformation within the EiBI theory in the literature, i.e., Sham {\it et al.} (2013)~\cite{Sham:2013cya}. There, they use the apparent EoS formulation of EiBI to simplify the star global properties calculations. Note that the authors~\cite{Sham:2013cya} focus on the case of  $\Lambda_c=0$.} Regarding some recent NS observational results, confronting NSs predicted by the EiBI theory with recent NS constraints~\cite{Landry:2020vaw,Jiang:2019rcw,Kumar:2019xgp,Lim:2018xne,Lattimer:2004nj,Breu:2016ufb} has not been performed yet. 
We also expect that setting $\lambda$ away from unity might increase $M$ while also decreasing $R$ to achieve relatively larger compactness. Therefore, in this study, we further investigate 
the role of the $\kappa$ and $\Lambda_c$ interplay 
in the predicted NS properties, such as mass, radius, \textcolor{black}{moment of inertia}, and tidal deformation. Then, we relate our results with the question of the tension between nuclear physics and NS property prediction, including the possibility that the secondary object of GW190814 is indeed a static or slow rotating NS. Here, we use the recent G3 RMF parameter set from Ref.~\cite{Kumara:2017bti} with hyperons as a representation of EoS and consider a speed-of-sound restriction at high densities when generating NS EoSs.

This paper is organized as follows: In Section~\ref{rmf}, we discuss the EoSs predicted by the RMF model in more detail. In Section~\ref{eibi}, we discuss the theoretical aspect of the EiBI theory, including the formulation of moments of inertia and tidal deformation in Subsections~\ref{momin} and~\ref{tidal}, respectively. In Section~\ref{result}, we show our numerical results and the corresponding discussions. Finally, in Section~\ref{concl}, we present the conclusions.

\section{EQUATION OF STATE OF NS MATTER }\label{rmf}

A summary of the RMF model description and the corresponding nuclear matter and NS matter predictions by some selected RMF parameter sets are presented in this section. Here, we show the reason of using the G3 RMF parameter set to describe the EoS of the core of NSs. In this section, we also show the reason to investigate the NS property predictions of the EiBI gravity theory.

The RMF Lagrangian density can be expressed as~\cite{Agrawal:2012rx}
\be
{\mathcal{L}} = {\mathcal{L}}_{B}  + 
{\mathcal{L}}_{BM}  + {\mathcal{L}}_{M} +{\mathcal{L}}_{L},  
\label{Lag}
\ee
where the free Lagrangian density for baryons ($B$ = $N$, $\Lambda$, $\Sigma$, $\Xi$) is
\be
{\mathcal{L}}_{B}=\sum_{B}\overline{\Psi}_B[i\gamma^{\mu}\partial_{\mu}-M_B]\Psi_B,
\label{eq:baryon}
\ee
where $M_B$ is the baryon mass and the Lagrangian density for meson-baryon couplings is given by
\bea                   
{\mathcal{L}}_{BM}&=& \sum_{B}\overline{\Psi}_B[g_{\sigma B} \sigma -\gamma_\mu g_{\omega B} \omega^\mu\nonumber\\&-&\frac{1}{2}\gamma_\mu g_{\rho B}\bm{\tau}_B\cdot \bm{\rho}^\mu -\gamma_\mu g_{\phi B}\phi ^\mu ]\Psi_B,
\label{eq:BM}
\eea
where the non-strange mesons that are coupled to all baryons are $\sigma$, $\omega$, and $\rho$. However, the hidden-strangeness meson $\phi$ is only coupled to hyperons ($H$ = $\Lambda$, $\Sigma$, $\Xi$). The free and self-interaction meson Lagrangian density can be expressed as  
\bea
{\mathcal{L}}_{M}&=&\frac{1}{2}(\partial_{\mu}\sigma\partial^{\mu}\sigma-m_{\sigma}^2\sigma^2)+\frac{1}{2}(\partial_{\mu}\sigma^*\partial^{\mu}\sigma^*-m_{\sigma^*}^2\sigma^{*2})\nonumber\\ &-&\frac{1}{4}\omega_{\mu\nu}\omega^{\mu\nu}+\frac{1}{2}m_{\omega}^2\omega_{\mu}\omega^{\mu}-\frac{1}{4}\phi_{\mu\nu}\phi^{\mu\nu}+\frac{1}{2}m_{\phi}^2\phi_{\mu}\phi^{\mu}\nonumber\\ &-&\frac{1}{4}\bm{\rho}_{\mu\nu}\cdot\bm{\rho}^{\mu\nu}+\frac{1}{2}m_{\rho}^2\bm{\rho}_{\mu}\cdot\bm{\rho}^{\mu}+{\mathcal{L}}^{NL}_{M},
\label{eq:meson}
\eea
The $\omega^{\mu\nu}$, $\phi^{\mu\nu}$ and $\bm{\rho}^{\mu\nu}$ are the meson tensor fields of the $\omega$, $\phi$, and $\rho$ mesons, which are defined as
$\omega^{\mu\nu}=\partial^{\mu}\omega^{\nu}-\partial^{\nu}\omega^{\mu}$, $\phi^{\mu\nu}=\partial^{\mu}\phi^{\nu}-\partial^{\nu}\phi^{\mu}$,
and $\bm{\rho}^{\mu\nu}=\partial^{\mu}\bm{\rho}^{\nu}-
\partial^{\nu}\bm{\rho}^{\mu}$. The explicit form of the Lagrangian density for meson self-interactions ${\mathcal{L}}^{NL}_{M}$ can be written as 
\bea
{\mathcal{L}}^{NL}_{M} &=& - \frac{\kappa_3 g_{\sigma N} m_{\sigma}^2}{6 m_{ N} } \sigma^{3}
- \frac{\kappa_4 g_{\sigma N}^2 m_{\sigma}^2}{24 m_{ N}^2 } \sigma^{4}+\frac{\zeta_0 g_{\omega N}^2}{24} 
{(\omega_{\mu}  \omega^{\mu})}^2\nonumber\\
&+& \frac{\eta_1 g_{\sigma N} m_{\omega}^2}{2 m_{ N} } \sigma \omega_{\mu}  \omega^{\mu}+\frac{\eta_2 g_{\sigma N}^2 m_{\omega}^2}{4 m_{ N}^2 }\sigma^{2} \omega_{\mu}  \omega^{\mu} 
\nonumber\\&+&\frac{\eta_{\rho} g_{\sigma N} m_{\rho}^2}{2 m_{ N} } \sigma
\bm{\rho}_{\mu} \cdot \bm{\rho}^{\mu} +\frac{\eta_{1\rho}g_{\sigma
		N}^2m_{\rho }^{2}}{4m_N^2} \sigma^2 \bm{\rho}_{\mu} \cdot \bm{\rho}^{\mu} \nonumber\\ &+&\frac{\eta_{2 \rho} g_{\omega N}^2 m_{\rho}^2}{4 m_{ N}^2 } \omega_{\mu}  \omega^{\mu} \bm{\rho}_{\nu} \cdot \bm{\rho}^{\nu}.
\label{eq:mesonNL}
\eea
Eq.~(\ref{eq:mesonNL}) includes the contribution from the standard RMF nonlinear self-interaction for $\sigma$ and $\omega$ mesons and additional cross-interaction terms for $\sigma$, $\omega$, and $\rho$ mesons. 
In RMF models, coupling constants and parameters in the Lagrangian density are determined by fitting the model predictions to finite nuclei and nuclear matter properties. The obtained parameter values depend on the chosen observables and their corresponding weights. The explicit value of the corresponding parameters of the RMF parameter sets used in this work can be found in Refs.\cite{Kumara:2017bti,Agrawal:2012rx,Shen:2020,Tolos:2017}.

\begin{figure}
	\centering
	\includegraphics[width=0.6\linewidth]{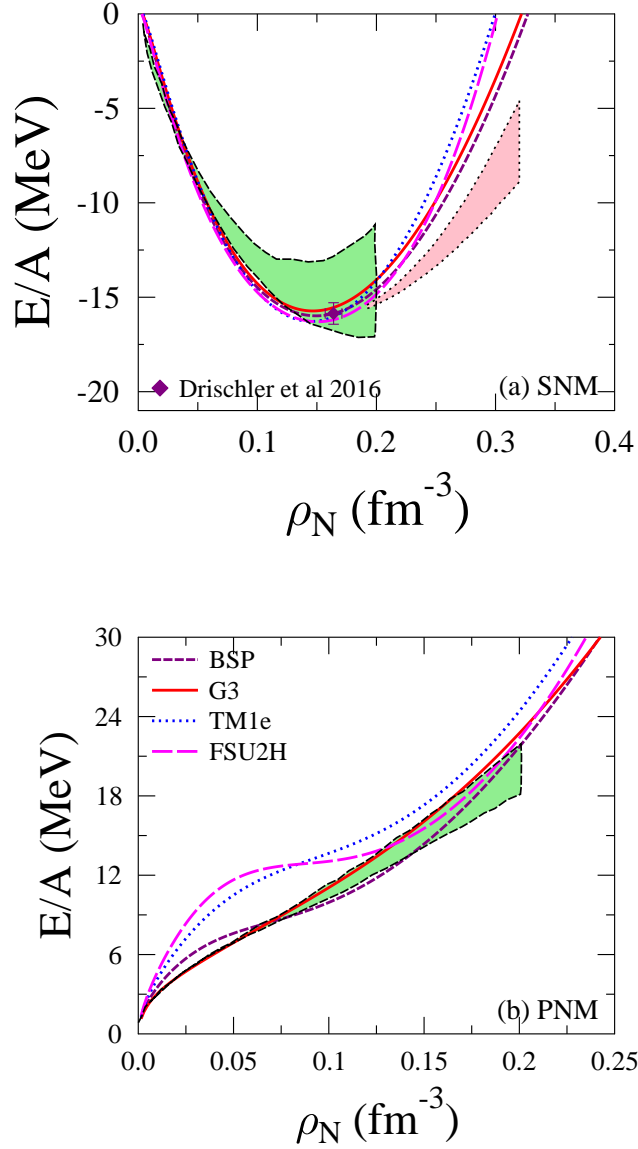}
	\caption{Binding energy predicted by G3, BSP, TM1e, and FSUH RMF parameter sets on SNM (a) and PNM (b). The light green-shaded area represents the chiral effective theory results taken from Ref.~\cite{Drischler2016}, whereas the pink-shaded area represents a constraint imposed by the SNM binding energy extracted from the FOPI experimental data~\cite{FOPI2016}. For comparison, the SNM binding energy at the saturation value \textcolor{black}{($\rho_0 \approx 0.16$ fm$^{-3}$)} from Ref.~\cite{Drischler2016} is also shown.}
	\label{fig:bindingenergyNM}
\end{figure}

\begin{figure}
	\centering
	\includegraphics[width=0.6\linewidth]{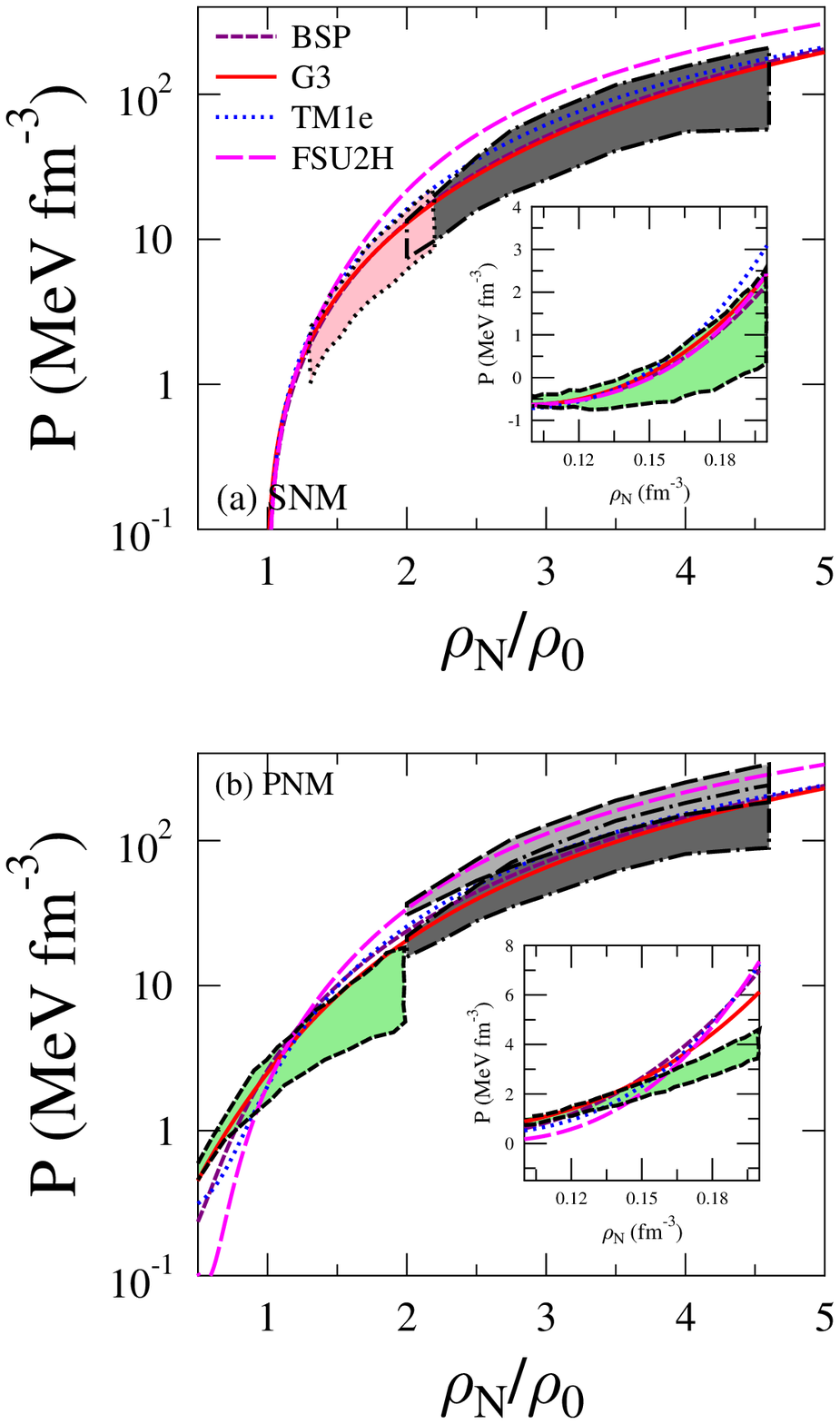}
	\caption{EoS represented by pressure as a function of the ratio nucleon density to the saturation density ($\rho_N$/$\rho_0$) in (a) SNM and (b) PNM, respectively. The results are calculated using the G3, BSP, TM1e, and FSUH RMF parameter sets. The gray-shaded areas are the results extracted from the heavy-ion experimental data~\cite{Daniel}. By contrast, the pink-shaded area in (a) is extracted from the FOPI experimental data~\cite{FOPI2016}, and the green-shaded areas in panels (a) and (b) are the theoretical binding energy for PNM at low densities obtained from the chiral effective field theory calculations~\cite{Drischler2016,Tew2018}. }
	\label{fig:EOSNM}
      \end{figure}

The contribution of EoSs in the nucleon sector is relatively established because the RMF parameter constraints in this sector are relatively tight. In Figs.~\ref{fig:bindingenergyNM} and \ref{fig:EOSNM} we show the binding energies and EoSs of symmetric nuclear matter (SNM) and pure neutron matter (PNM) predicted by G3~\cite{Kumara:2017bti}, BSP~\cite{Rahmansyah:2020gar,Agrawal:2012rx,Sulaksono:2012ny}, TM1e~\cite{Shen:2020}, and FSUH~\cite{Tolos:2017} parameter sets. The results are compared to those extracted from experimental data~\cite{FOPI2016,Daniel} and those obtained from the chiral effective field theory calculations~\cite{Drischler2016,Tew2018}. In general, all parameter sets are compatible with experimental data, but at low densities the G3 results are more compatible with those obtained from the chiral effective field theory calculations than those from other parameter sets used in this work. \textcolor{black}{However, it can be observed that binding energies predicted by all RMF parameter sets that we use here are not too compatible with the binding energy constraint from FOPI for $\rho_N \le 2 \rho_0$. On the other hand, the EoS of the G3 parameter set is compatible with the EoS constraint from FOPI.\footnote{Note that we made the following attempts to check whether our results are  ``correct'' or not. First, to avoid false data extractions; we have already rechecked and compared the extracted binding energy and the EoS data with the ones from the paper~\cite{FOPI2016}. They are now precisely matched. Second, we have also recheck the RMF binding energy and EOS subroutines in our code. It seems that we have made no mistakes because, in the RMF code, we solved the equations self-consistently. If we make a mistake in one quantity, the error will truncate to all quantities because they are strongly correlated. Third, we compared the results with other people's published calculation results using RMF models~\cite{Kumara:2017bti,Agrawal:2012rx,Shen:2020,Tolos:2017}, and the results are pretty compatible.} Therefore, we only can argue that on the RMF calculation side, the binding energy and EoS results are consistent.} \textcolor{black}{To this end}, in this work, we decided to use the G3 parameter set as the representative parameter set to study NS properties.   

Generally, hyperons and other exotic particle coupling constants are experimentally difficult to constrain. Therefore, the contribution of EoSs in the hyperon sector is uncertain. The inclusion of hyperons and other exotic particles tends to soften the corresponding EoS of the NS core. Therefore, the corresponding predicted maximum mass is always smaller than that obtained without hyperons and other exotics\footnote{See, for example, Ref.~\cite{Tolos:2017} and the references therein for related hyperon puzzle discussion.}. Following Ref.~\cite{Tolos:2017}, here, we take the SU(3) prescription and experimental value of potential depths at the nuclear matter saturation density to determine the hyperon coupling constants while neglecting the contribution from other exotics. The SU(3) prescription yields a relatively stiffer EoS compared to that of SU(6)~\cite{Rahmansyah:2020gar}. For leptons, we use the free Lagrangian density. To describe the NS's crusts, we use the inner and outer crust EoSs based on the Hartree-Fock Thomas-Fermi model used by Miyatsu {\it et al.} (2012)~\cite{MYN2013}\footnote{See Ref. \cite{Rahmansyah:2020gar} and the references therein for the detailed discussion about the uncertainty that should be paid using these crust EoSs.}. The NS matter is assumed in $\beta$-stability. Therefore, the potential chemical balance, charge neutrality, and baryon density conservation conditions can be used to determine the constituents' composition in NSs. Here, we also generate the EoSs of NSs constrained by the speed of sound bound at high densities $v_s$ $\le$ $c$/$\sqrt{3}$ (G3 WoutHSS and G3 WHSS), where $c$ is the speed of light\footnote{See Ref.~\cite{Margaritis2020} and references therein on the recent progress of speed-of-sound constraints on NS discussions.}.       
        \begin{figure}
	\centering
	\includegraphics[width=0.5\linewidth]{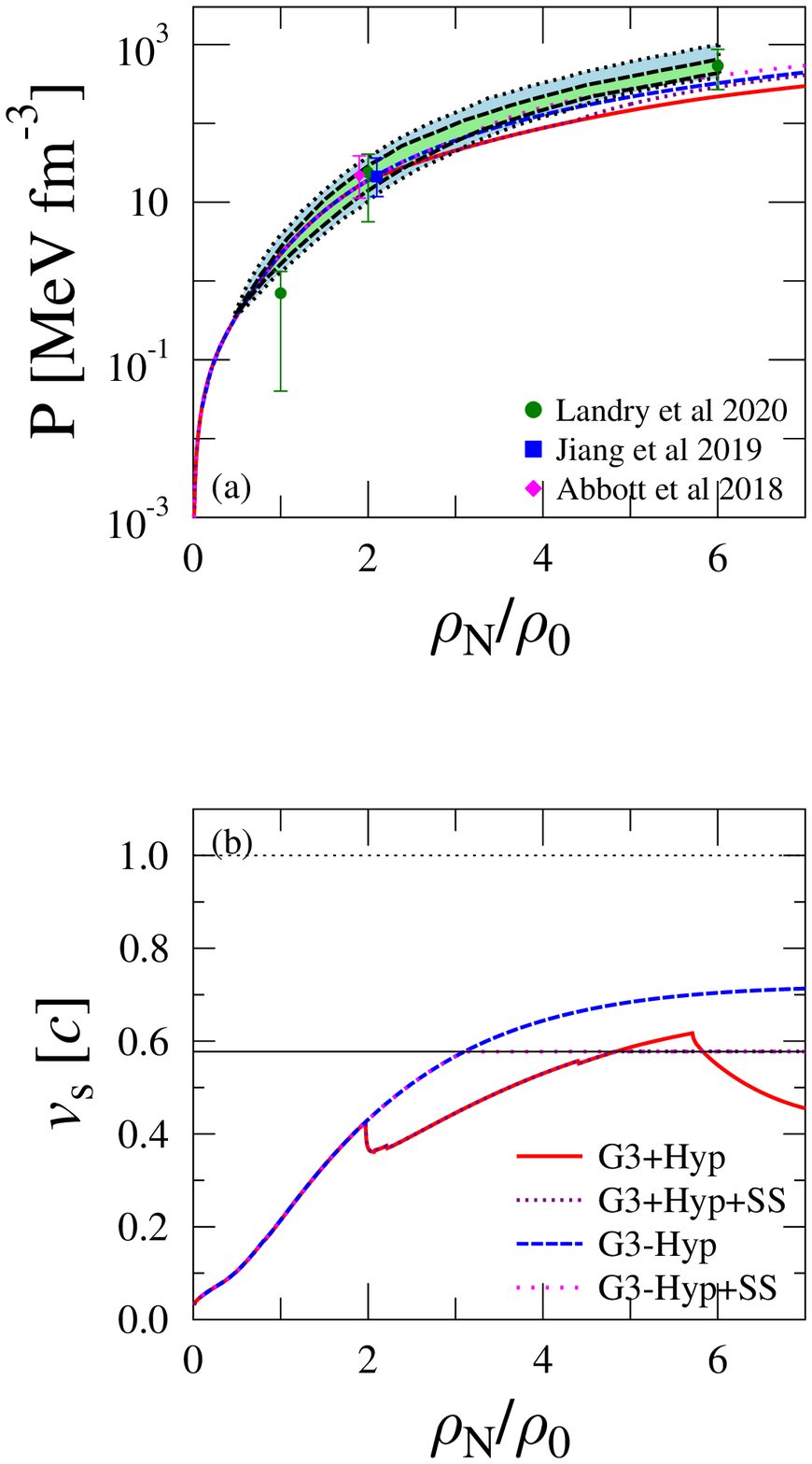}
	\caption{EoS without a hyperon (G3-Hyp) and with a hyperon (G3+Hyp) is calculated using the G3 parameter set. We also show the EoSs if the speed of sound at high densities is constrained by $v_s$ $\le$ $c$/$\sqrt{3}$ (G3-Hyp+SS and G3+Hyp+SS). In (a), we show pressure as a function of the ratio of $\rho_N$ to $\rho_0$, and in (b), we show the speed of sound as the function of the ratio of density to saturation density. For comparison, some NS EoS constraints are given: the light-blue and light-green shaded areas from GW170817~\cite{Abbott2018} and the data points for particular densities taken from the GW170817~\cite{Abbott2018} results, recent non-parametric analysis~\cite{Landry:2020vaw}, and those from the joint of PSR J0030+0451, GW170817, and the nuclear data analysis from Ref.~\cite{Jiang:2019rcw}.}
	\label{fig:EOSNSG3}
      \end{figure}      

\begin{figure}
	\centering
	\includegraphics[width=0.6\linewidth]{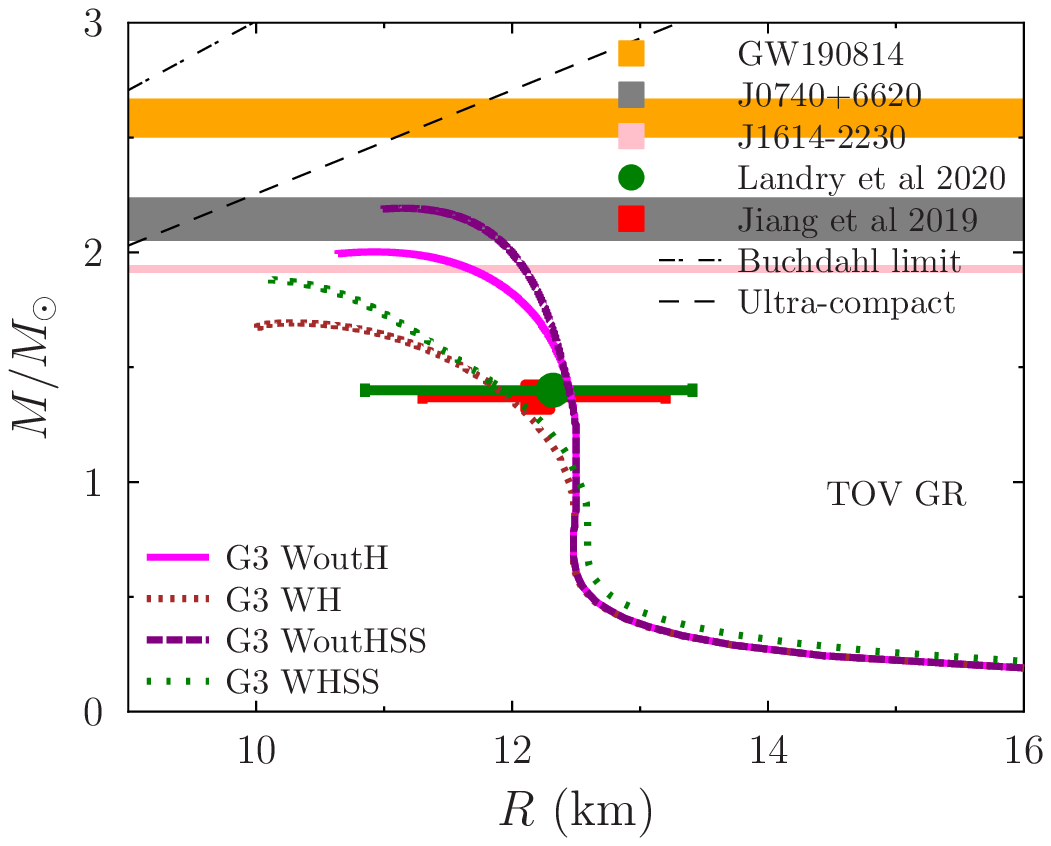}
	\caption{Mass-radius relation predicted by the G3 parameter set for the case matter without a hyperon (WoutH) and with a hyperon (WH) using the GR framework. We also show the relations if the sound speed is constrained at high densities (WoutHSS and WHSS). }
	\label{fig:plotvareos}
\end{figure}

The NS EoSs and the corresponding speed of sound for the case matters without a hyperon (G3 WoutH) and with a hyperon (G3 WH) and those with the speed-of-sound constraint (G3 WoutHSS and G3 WHSS) are shown in Fig.~\ref{fig:EOSNSG3}. Simultaneously, the corresponding mass-radius relations within GR are shown in Fig.~\ref{fig:plotvareos}. In Fig.~\ref{fig:EOSNSG3}, the recent EoS constraints~\cite{Abbott2018,Landry:2020vaw,Jiang:2019rcw} are more compatible with the NS EoS without hyperons (G3 WoutH), and the sound of the speed restricted at high-density treatments slightly increases the stiffness of the corresponding EoSs. However, for EoSs with hyperons (G3 WoutHSS), the effect is insufficient to reach the EoS constraint from GW170817 at high densities. The impacts of hyperons and the speed-of-sound constraint on the mass-radius relation within the GR theory are shown in Fig.~\ref{fig:plotvareos}. The data from the GW190814 event were obtained from Ref.~\cite{Abbott:2020khf}, and the others were from Landry \textcolor{black}{\it et al.} (2020)~\cite{Landry:2020vaw} and Jiang \textcolor{black}{\it et al.} (2019)~\cite{Jiang:2019rcw}. We also show the results from the pulsar-binary system analysis (PSR) J1614-2230 by Refs.~\cite{Demorest:2010bx,Fonseca:2016tux,Arzoumanian:2017puf} and J0740+6620 by Refs.~\cite{Arzoumanian:2017puf,Cromartie:2019kug}. The boundaries where the ultra-compact limit and Buchdahl limit are located are also shown. The figure clearly shows that hyperon's contribution lowers the mass, and constraining sound speed increases the mass. All EoSs are compatible with radius constraints from Refs~\cite{Landry:2020vaw,Jiang:2019rcw}. However, hyperons' contribution on EoS lowers the NS maximum mass significantly below the 2 $M_\odot$ pulsar mass constraints. Considering the requirement that the speed of sound should be less than $v_s$ $\le$ $c$/$\sqrt{3}$ in EoSs slightly increases the NS maximum mass. However, the corresponding maximum mass is still less than the 2 $M_\odot$ pulsar mass constraint. If the second object with mass 2.6 $M_\odot$ detected by the LIGO collaboration (GW 190814) is a non-rotating NS, then all maximum masses predicted by all EoSs used in this work are less than this constraint. Recent studies~\cite{Fattoyev:2020cws,Das:2020dcq} have shown that if the hyperons are excluded in the NS matter, then 2.6 $M_\odot$ and canonical NS radius constraints can simultaneously be satisfied using a particular RMF EoS (Big Apple), which is compatible with finite nuclei and nuclear matter constraints. However, the corresponding nuclear matter EoS prediction is not compatible with those obtained from heavy-ion collision constraints~\cite{Daniel}. 
Therefore, Fattoyev {\it et al.} (Ref.~\cite{Fattoyev:2020cws}) concluded that the 2.6 $M_\odot$ compact object is not likely an NS within a GR.
To this end, it is worth noting that the EiBI gravity can have maximum NSs with an acceptable EoS larger than 2 $M_\odot$ without reaching the Buchdahl limit~\cite{Qauli:2016vza}. However, according to the EiBI theory, as the maximum mass increases, the radius also increases. Therefore, in the next sections, we will systematically examine the tension between relatively small recent radius constraints and recent considerable maximum mass constraints within the EiBI theory.   

\section{EIBI THEORY}\label{eibi}

In this section, we briefly review the EiBI theory and discuss the moment of inertia and tidal deformation derivation within it. We start by reviewing the formulas following the treatment proposed in Refs.~\cite{Harko:2013wka,Qauli:2016vza}. The EiBI theory has the following equations of motion:
\begin{eqnarray}
    q^{\mu\nu}&=&\tau \left(\lambda g^{\mu\nu}-8\pi G\kappa T^{\mu\nu}(\bm{g})\right),\label{eq:0}\\
    q_{\mu\nu}&=&g_{\mu\nu}+\kappa R_{\mu\nu}(\bm{q}),\label{eq:0.1}
\end{eqnarray}
where $\tau={\sqrt{g}}/{\sqrt{q}},~q=-\det(q_{\mu\nu}),~g=-\det(g_{\mu\nu}).$ (It is a usual practice to use $\tau=\sqrt{\det(g_{\sigma\nu})\det(q^{\mu\sigma})}=[\det(\lambda\delta^\mu_\nu-8\pi G\kappa T^\mu_\nu)]^{-1/2}$.)
Here, $\lambda$ and $\kappa$ are the parameters of EiBI, different from functions $\bar{\lambda}(r)$ and $\tilde{\kappa}(r)$, which we shall define below. $\kappa$ has dimension (length)$^2$, and $\lambda$ is dimensionless. The cosmological constant $\Lambda_{c}$ is related to both of them by
\begin{equation}
    \lambda=\kappa\Lambda_c+1.
\end{equation}
The arguments in $R_{\mu\nu}$ and $T^{\mu\nu}$ are different because to raise or lower the tensor indices, each using a different metric, i.e., the apparent metric $q_{\mu\nu}$ and physical metric $g_{\mu\nu}$, respectively. These equations are products of the EiBI action:
\begin{equation}
    S={1\over 8\pi G\kappa}\int_{(\mathcal{M},\bm{g})}d^4x~\left[\sqrt{-\det(g_{\mu\nu}+\kappa R_{\mu\nu}(\Gamma))}-\lambda\sqrt{-\det(g_{\mu\nu})}\right],
\end{equation}
where the Palatini formalism has been used, i.e., the Ricci tensor is dependent not on the physical metric but on the connection $\Gamma^\alpha_{\beta\gamma}$, which is dependent on the apparent metric.
\begin{equation}
    \Gamma^\alpha_{\beta\gamma}={1\over 2}q^{\alpha\sigma}\left(\partial_\gamma q_{\beta\sigma}+\partial_\beta q_{\gamma\sigma}-\partial_\sigma q_{\beta\gamma}\right).
\end{equation}
To make them similar to the Einstein field equation (EFE), we can manipulate them into
\begin{eqnarray}
    &&R^\mu_\nu(\bm{q})-\frac{1}{2}R^\sigma_\sigma \delta^\mu_\nu(\bm{q})=8\pi G T^\mu_{\mathrm{eff}~\nu} (\bm{g}),\\
    &&T^\mu_{\mathrm{eff}~\nu} (\bm{g})=\tau T^\mu_{\nu} (\bm{g})-\left[\frac{\tau T^\sigma_\sigma (\bm{g})}{2}+\frac{1-\tau\lambda}{8\pi G\kappa}\right]\delta^\mu_\nu.
\end{eqnarray}
For brevity, we shall not write the arguments of the Ricci tensor $R_{\mu\nu}$ and stress tensor $T_{\mu\nu}$.

\subsection{Vacuum solution}
In this paper, we will discuss the effect of the nonzero cosmological constant $\Lambda_c$ on the EiBI theory. Before we proceed, it is necessary to discuss the vacuum solution. Suppose we have the apparent metric and physical metric in the following static and spherically symmetric form:
\begin{eqnarray}
    \bm{q}=q_{\mu\nu}dx^\mu dx^\nu&=&-C^2(r) dt^2+D^2(r) dr^2 +r^2 [d\theta^2+\sin^2\theta d\varphi^2],\\
    \bm{g}=g_{\mu\nu}dx^\mu dx^\nu&=&-A^2(r) dt^2+B^2(r) dr^2 +F^2(r) [d\theta^2+\sin^2\theta d\varphi^2],
\end{eqnarray}
and we have no matter at all (i.e., $T^\mu_\nu=0$). From Eq. \eqref{eq:0}, we have
\begin{eqnarray}
F^2&=&r^2/\lambda,\\
A^2&=&C^2/\lambda,\\
B^2&=&D^2/\lambda.
\end{eqnarray}
Substituting these into Eq. \eqref{eq:0.1}, we have
\begin{eqnarray}
C''&=&\frac{DC-D^3C+rCD'+rDC'+r^2D'C'}{r^2D},\\
{C'\over C}&=&-{1\over r}+\left({1\over r}-{r\over \kappa}+{r\over \kappa\lambda}\right)D^2+{D'\over D},\\
{D'\over D}&=&{1\over 2r}+\left(-{1\over 2r}+{r\over 2\kappa}-{r\over 2\kappa\lambda}\right)D^2,
\end{eqnarray}
where the primes denote the differentiation with respect to $r$. The solutions that satisfy these equations are
\begin{equation}
C^2=D^{-2}=1-{2GM\over r}-{\Lambda_c r^2\over 3\lambda}.
\end{equation}
The apparent metric without the presence of matter indicates an ``apparent'' Minkowski-de Sitter space with the cosmological constant $\Lambda_c/\lambda$. This factor will be crucial for our metric ansatz with the presence of an ideal isotropic fluid so that we can obtain suitable equations of motion.

\subsection{Moment of Inertia}
\label{momin}

In this subsection, first, we set the apparent and physical metrics in the following forms:
\begin{eqnarray}
    \bm{q}=q_{\mu\nu}dx^\mu dx^\nu&=&-e^{\beta(r)} dt^2+e^{\alpha(r)} dr^2 +r^2 [d\theta^2+\sin^2\theta(d\varphi-\omega(r) dt)^2]    +\mathcal{O}(\Omega^2),\\
    \bm{g}=g_{\mu\nu}dx^\mu dx^\nu&=&-e^{\nu(r)} dt^2+e^{\bar{\lambda}(r)} dr^2 +d(r) [d\theta^2+\sin^2\theta(d\varphi-v(r) dt)^2]    +\mathcal{O}(\Omega^2).
\end{eqnarray}
Our apparent metric is the Hartle-Thorne metric~\cite{Hartle:1967he} and the physical metric is its generalization. Both describe a spherically symmetric massive body with radius $R$ with angular momentum $\omega\sim\Omega$ (and $v\sim\Omega_\mathrm{phy}$) as $r\to R$.  
If $\Omega_k$ is defined as the Kepler angular velocity, then it is assumed that $\Omega/\Omega_k\ll 1$ and $\Omega_\mathrm{phy}/\Omega_k \ll 1$. This is known in the literature as a slow rotating approximation~\cite{Hartle:1967he}. 
We construct the physical metric whose 2-sphere has radius $\sqrt{d(r)}$. The boundary conditions for the metrics is that both of them have the same exterior region, so both should coincide at $r\geq R$.

The massive body is assumed to be an ideal fluid, such that:
\begin{eqnarray}
    T^\mu_{\nu}&=&\left[\epsilon+p\right]u^\mu u_\nu+p \delta^\mu_\nu,\\
    u^t&=&[-(g_{tt}+2\Omega_\mathrm{phy} g_{t\varphi}+\Omega_\mathrm{phy}^2 g_{\varphi\varphi})]^{-1/2},\\
    u^\varphi&=&\Omega_\mathrm{phy} u^t,~ u^r=u^\theta=0.
\end{eqnarray}
In its explicit form, the components in the physical stress tensor are as follows:
\begin{eqnarray}
    &&T^t_t=-\epsilon,~T^r_r=T^\theta_\theta=T^\varphi_\varphi=p,\\
    &&T^t_\varphi=(\epsilon+p)(\Omega_\mathrm{phy}-v)e^{-\nu} d\sin^2\theta,\\
    &&T^\varphi_t=-(\epsilon+p)\Omega_\mathrm{phy}.
\end{eqnarray}
Then, after neglecting $\mathcal{O}(\Omega_\mathrm{phy})$, the explicit form of $\tau$ becomes 
\begin{eqnarray}
    \tau&=&1/(ab^3),\\
    a&=&\sqrt{\lambda+8\pi G\kappa\epsilon},\\
    b&=&\sqrt{\lambda-8\pi G\kappa p}.
\end{eqnarray}

From Eq. \eqref{eq:0}, we have the diagonal components from both metrics related by
\begin{equation}
    e^{\nu}=e^\beta a/b^3,~ e^{\bar{\lambda}}=e^\alpha /(ab),~ d=r^2/(ab).
\end{equation}
We assume that the effective stress tensor also has a similar form as the physical stress tensor but with additional subscript ``eff'', i.e.,
\begin{eqnarray}
T^t_{\mathrm{eff}~t}&=&-\epsilon_\mathrm{eff},\\
T^r_{\mathrm{eff}~r}&=&p_\mathrm{eff}=T^\theta_{\mathrm{eff}~\theta}=T^\varphi_{\mathrm{eff}~\varphi},\\
T^t_{\mathrm{eff}~\varphi}&=&(\epsilon_\mathrm{eff}+p_\mathrm{eff})(\Omega-\omega)e^{-\beta(r)}r^2\sin^2\theta.
\end{eqnarray}
Then, we obtain
\begin{eqnarray}
    \epsilon_\mathrm{eff}&=&\frac{a^2-3b^2+2ab^3}{16\pi G\kappa ab^3},\label{eq:edenapp}\\
    p_\mathrm{eff}&=&\frac{a^2+b^2-2ab^3}{16\pi G\kappa ab^3},\label{eq:pressapp}\\
    (\Omega-\omega)
    &=&(\Omega_\mathrm{phy}-v)\frac{b^2}{a^2}.
\end{eqnarray}
The last equation was derived from $T^t_{\mathrm{eff}~\varphi}={T^t_{~\varphi}}/{ab^3}.$
Because $(\Omega-\omega)\sim(\Omega_\mathrm{phy}-v)$ at $r\to R$, we can demand constants $\Omega$ and $\Omega_\mathrm{phy}$ to satisfy 
\begin{equation}
    \Omega=\Omega_\mathrm{phy}.\label{eq:OmegaPhy}
\end{equation}
This constraint is actually justified because it came from $T^\varphi_{\mathrm{eff}~t}={T^\varphi_{~t}}/{abc^2}$.

The components of the Ricci tensor can be obtained in a straightforward manner. Defining mass inside one of the metric function
\begin{equation}
    e^{-\alpha}=1-{2Gm(r)\over r}-{\Lambda_cr^2\over 3\lambda},\label{eq:3}
\end{equation}
we obtain
\begin{eqnarray}
    m'(r)&=&{r^2\over 4G\kappa}\left({2\over \lambda}-{3\over ab}+{a\over b^3}\right),\label{eq:m'}\\
    \beta'(r)&=&-2p'(r) \left[
    2\pi G\kappa\left(
    {3\over b^2}+{1\over a^2}{d\epsilon\over dp}
    \right)+{1\over \epsilon+p}
    \right]
    ,\label{eq:4}\\
    p'(r)&=&-{1\over 4\pi G\kappa}\left[ 
    {r\over 2\kappa}\left(
    {1\over ab}+{a\over b^3}-2
    \right) 
    +{2Gm\over r^2}+{\Lambda_c r\over 3\lambda}
    \right]\nonumber\\
    &&\times \left(
    {4\over a^2-b^2}+{3\over b^2}+{1\over a^2}{d\epsilon\over dp}
    \right)^{-1}
    \left(1-{2Gm(r)\over r}-{\Lambda_cr^2\over 3\lambda}\right)^{-1}, \label{eq:p'}
\end{eqnarray}
from $tt$, the $rr$ components of the EFE, and the (contracted) Bianchi identity $\nabla_\mu T^\mu_{\mathrm{eff}~r}=0$, respectively.
The boundary conditions are $m(0)=0, m(R)=M,~ p(R)=0, and ~ \beta(R)=\ln\left(1-{2GM/R}-{\Lambda_cR^2/ (3\lambda)}\right)$. Clearly, transforming $\beta\to\beta+k$ ($k$ a constant) does not change the equation of motion, so we can easily obtain $\beta(0)=\beta_{0,\text{old}}$ from the arbitrary value of $\beta_0$, run the code to obtain $\beta_R=\beta(R)$, and then run again the code for the second time from the initial value:
\begin{equation}
\beta_{0,\text{new}}=\beta_{0,\text{old}}-\left[
\beta_R-\ln\left(1-{2GM/R}-{\Lambda_cR^2/ (3\lambda)}\right)
\right].\label{eq:betacorr}
\end{equation}

To obtain an equation for the moment of inertia, we calculate the equation of motion of $\omega$ from the $t\varphi$ component of the EFE. To have this, we use the following formula:
\begin{equation}
    R^t_\varphi={1\over \sqrt{-\det(q_{\alpha\beta})}}\partial_\mu
    \left(\sqrt{-\det(q_{\alpha\beta})}\Gamma^\mu_{\nu\varphi}q^{t\nu}\right).\label{eq:2.0}
\end{equation}
With Eq. \eqref{eq:2.0} and ignoring $\mathcal{O}(\omega^2)$, we obtain
\begin{equation}
    R^t_\varphi=-\frac{e^{-(\beta+\alpha)/2}}{2r^2\sin\theta}\partial_r\left(e^{-(\beta+\alpha)/2} r^4 \sin^3\theta \partial_r\omega\right).\label{eq:Rtphi}
\end{equation}
From before, we have $T^t_{\mathrm{eff}~\varphi}=(\epsilon_\mathrm{eff}+p_\mathrm{eff})(\Omega-\omega)e^{-\beta(r)}r^2\sin^2\theta$.
Now, we define $\tilde{\omega}=(\Omega-\omega)/\Omega$.
Then, $R^t_\varphi/(8\pi G)=T^t_{\mathrm{eff}~\varphi}$ becomes
\begin{eqnarray}
    \partial_r\left(e^{-(\beta+\alpha)/2} r^4 \partial_r\tilde{\omega}\right)
    =16\pi G r^4 (\epsilon_\mathrm{eff}+p_\mathrm{eff})e^{(\alpha-\beta)/2}\tilde{\omega}.\label{eq:5}
\end{eqnarray}
In the exterior region, the right-hand side vanishes, so
\begin{equation}
    \tilde{\omega}(r\geq R)=1-\frac{2GI}{r^3},
\end{equation}
which is the boundary condition to calculate the moment of inertia $I$.
From Eq. \eqref{eq:5}, we have
\begin{eqnarray}
    \tilde{\omega}'(r)&=&\frac{6 e^{\beta/2}}{r^4(1-2Gm/r-\Lambda_c r^2/(3\lambda))^{1/2}}\tilde{\kappa},\label{eq:7}\\
    \tilde{\kappa}'(r)&=&\frac{8\pi Gr^4}{3}  \frac{(\epsilon_\mathrm{eff}+p_\mathrm{eff})e^{-\beta/2}}{(1-2Gm/r-\Lambda_c r^2/(3\lambda))^{1/2}}\tilde{\omega},\label{eq:8}
\end{eqnarray}
whose boundary conditions are $\tilde{\omega}(R)=1-2GI/R^3,$ and $\tilde{\kappa}(R)=GI.$
Because the boundary condition at the center is unknown, we pay attention on Eqs. \eqref{eq:7} and \eqref{eq:8}. Notice that both are invariant from replacing $\tilde{\omega}(r)\to \zeta \tilde{\omega}(r)$ and $\tilde{\kappa}(r)\to \zeta \tilde{\kappa}(r)$. Suppose that the results of the numerical calculations give us $\tilde{\omega}(R)= (1-2GI/R^3)/\zeta$ with $\tilde{\kappa}(R)= GI/\zeta,$ and with $\zeta$ as a constant, from initial values $\tilde{\omega}(0)= \tilde{\omega}_0$ and $\tilde{\kappa}(0)= \tilde{\kappa}_0$. Then, to satisfy both boundary conditions, we can set the initial values to be $\tilde{\omega}(0)= \tilde{\omega}_0\zeta$ and $\tilde{\kappa}(0)= \tilde{\kappa}_0\zeta$ with
\begin{equation}
    \zeta={1\over \tilde{\omega}(R)+2\tilde{\kappa}(R)/R^3}. \label{eq:zeta}
\end{equation}
However, recalculating is unnecessary because we already obtained the moment of inertia $I$ from $\tilde{\kappa}(R)$ by $I=\tilde{\kappa}(R) \zeta /G$. Following Refs. \cite{Pani:2012qb,Pani:2011mg}, we have $v(R)=\omega(R)=2I\Omega/R^2$ from the boundary condition. Thus, we already obtain $I$ as the physical moment of inertia.

The numerical procedure for the moment inertia is as follows: First, we calculate all $p'(r)$, $m'(r)$ and $\nu'(r)$ (Eqs.~\eqref{eq:p'}-\eqref{eq:m'}). We employ the Runge-Kutta 4th-order algorithm using a {\it FORTRAN77} code. The initial data at the center $r=r_c\to 0$ are $p(r_c)=p_c$, $m(r_c)=0$, and $\beta(r_c)=0$. We run the code up to $r=R$, where the pressure becomes zero $p(R)=0$. At this point, we obtain $R$, $m(R)=M$ and $\beta(R)=\beta_R$. Because in general the value of $\beta(R)$ is not equal to $\ln(1-2GM/R-\Lambda_c R^2/(3\lambda))$, we use a new initial value $\beta(r_c)=\beta_{0,\text{new}}$ using Eq.~\eqref{eq:betacorr}. Second, we calculate all $p'(r)$, $m'(r)$, $\nu'(r)$, $\tilde{\omega}'(r)$, and $\tilde{\kappa}'(r)$. The initial values
at $r=r_c$ are
$p(r_c)=p_c$, $m(r_c)=0$, $\beta(r_c)=\beta(r_c)_\text{new}$, and $\tilde{\omega}(r_c)=\tilde{\kappa}(r_c)=0$. The new results are $\tilde{\omega}(R)$ and $\tilde{\kappa}(R)$. The moment of inertia $I$ is determined by $I=\tilde{\kappa}(R) \zeta /G$ with $\zeta$ from Eq.~\eqref{eq:zeta}.

\subsection{Tidal Deformation}
\label{tidal}

In this subsection, we focus only on tidal deformation for the electric type. We start with the following unperturbed metrics:
\begin{eqnarray}
\eta_{\mu\nu}dx^\mu dx^\nu&=&
-e^{\nu(r)} dt^2 + e^{\bar{\lambda}(r)} dr^2
+ d(r) d\Omega^2,
\\
\zeta_{\mu\nu}dx^\mu dx^\nu&=&
-e^{\beta(r)} dt^2 + e^{\alpha(r)} dr^2
+ r^2 d\Omega^2,
\end{eqnarray}
where $d\Omega^2$ is the surface element of 2-sphere. Following the Regge-Wheeler metric, the perturbed metrics are
\begin{eqnarray}
    g_{\mu\nu}&=&\eta_{\mu\nu}+h_{\mu\nu},
    \\
    q_{\mu\nu}&=&\zeta_{\mu\nu}+f_{\mu\nu},
\end{eqnarray}
with
\begin{eqnarray}
h_{\mu\nu}&=&\begin{pmatrix}
    -H_0 e^\nu & H_1 & 0 & 0 \\
    H_1 & H_2 e^{\bar{\lambda}} & 0 & 0 \\
    0 & 0 & Kr^2 & 0 \\
    0 & 0 & 0 & Kr^2\sin^2\theta 
\end{pmatrix}
Y_{lm}(\theta,\phi)
,\\
f_{\mu\nu}&=&\begin{pmatrix}
    -F_0 e^\nu & F_1 & 0 & 0 \\
    F_1 & F_2 e^{\bar{\lambda}} & 0 & 0 \\
    0 & 0 & \bar{G}r^2 & 0 \\
    0 & 0 & 0 & \bar{G}r^2\sin^2\theta 
\end{pmatrix}
Y_{lm}(\theta,\phi)
.
\end{eqnarray}
All $H_0,~H_1,~H_2,~K,~F_0,~F_1,~F_2,$ and $\bar{G}$ are functions of $r$.

\subsubsection{Vacuum case}

To determine the Love number, first, we investigate in the vacuum case. From Eq. \eqref{eq:0} with $T^\mu_\nu=0$, one obtains
\begin{eqnarray}
d=r^2/\lambda,~e^\nu=e^\beta/\lambda,~e^{\bar{\lambda}}=e^\alpha/\lambda,~\\
H_0=-H_2=H, F_0=-F_2=F,~\\
H=F,~ K=\bar{G}/\lambda.
\end{eqnarray}
From evaluating Eq. \eqref{eq:0.1} in order, the following solutions and equations are obtained:
\begin{eqnarray}
e^\beta&=&e^{-\alpha}=1-{2GM\over r}-{\Lambda_c r^2\over 3\lambda},\\
F_1&\propto& e^{(\alpha-\beta)/2},\\
H_1&=&F_1\left(
{1\over \lambda}-{l(l+1)\kappa\over 2r^2}
\right),\\
F''&+&2e^\alpha\left(
{1\over r}-{\Lambda_c r\over \lambda}
\right) F' 
-2\left(
{1\over r^2}-e^\alpha{\Lambda_c \over \lambda}
\right) F
+e^\alpha {(l-1)(l+2)\over r^2}G=0.
\end{eqnarray}
After some algebraic manipulations, the last line above becomes
\begin{equation}
F''+\left[{1\over r}+e^\alpha\left(
{1\over r}-{\Lambda_c r\over \lambda}
\right)\right] F'
-\left[
e^a\left({l(l+1)\over r^2}
+{2\Lambda_c\over\lambda}\right)+\beta'^2
\right]F=0.\label{eq:Fvacuum}
\end{equation}

Now, we come to the tricky part. To calculate the tidal deformation, the calculation is performed at $r\to\infty$, but de Sitter space is not asymptotically flat. 
To remedy this limitation, we assume that $|\Lambda_c|$ is sufficiently small such that $|(GM)^2\Lambda_c/\lambda|\ll 1$.
Thus, the solution $F$ is assumed to have the following form:
\begin{equation}
F(x)=\sum_{i=0}^{\infty}F_i(x) \varepsilon^i,~~  
\varepsilon={G^2M^2\Lambda_c\over \lambda},~~
x={r\over GM}-1.
\end{equation}
Assuming that the series rapidly converges, we consider the series only up to the first order:
\begin{equation}
F(x)=F_0(x)+\varepsilon F_1(x).\label{eq:assumesmallLam}
\end{equation}
From $\Lambda_c=0$, we have the following usual solution:
\begin{equation}
F_0(x)=C_{1,l} Q^2_l(x)+C_{2,l} P^2_l(x),
\end{equation}
where $Q^2_l$ and $P^2_l$ are the associated Legendre polynomials of the second and first kinds, respectively. The constants $C_{1,l}$ and $C_{2,l}$ will be determined later. Substituting this into Eq. \eqref{eq:Fvacuum}, we obtain
\begin{equation}
(1-x^2)\left({d^2F_1(x)\over dx^2}+D(x)\right)
-2x{dF_1(x)\over dx}
+\left(l(l+1)-{4\over 1-x^2}\right)F_1(x)=0,
\end{equation}
with
\begin{equation}
D(x)=-{1\over 3}\left({x+1\over x-1}\right)^2\left[
2(x-2){dF_0(x)\over dx}
+\left(
l(l+1)-{6x^2-20x+22\over 1-x^2}
\right)F_0(x)
\right].
\end{equation}
Then, we solve this equation with $l=2,3,4$ case by case. 

In general, the obtained solution has the following form: 
\begin{eqnarray}
	F_1(x)=C_{3,l} Q^2_l(x) 
	+C_{4,l} P^2_l(x)
	+C_{2,l} S^2_l(x)
	+C_{1,l} T^2_l(x)
\end{eqnarray}
with
\begin{eqnarray}
	S^2_l(x)&=&
	{f_{1,l}(x)\over x^2-1}+(x^2-1)f_{2,l}(x),\\
	T^2_l(x)&=&
	{f_{3,l}(x)\over(x+1)(x-1)^2}+{f_{4,l}(x)\over x+1}\ln(x-1)
	\nonumber\\
	&&+f_{5,l}(x)\left(
	{x+1\over x-1}
	\right)\ln(x+1)
	+(x^2-1)f_{6,l}(x).
\end{eqnarray}
For $l=2$, we have
\begin{eqnarray}
f_{1,2}&=&
\frac{8 x^6}{7}+6 x^5-\frac{x^4}{7}-\frac{59 x^3}{4}-\frac{46 x^2}{7}+\frac{21 x}{4}+1
,\\
f_{2,2}&=&
\frac{3}{56} (113 \log (x-1)+15 \log (x+1))
,\\
f_{3,2}&=&
-\frac{8 x^6}{7}+\frac{389 x^5}{56}+\frac{2057 x^4}{168}-\frac{1987 x^3}{84}-\frac{1469 x^2}{84}+\frac{1357 x}{56}+\frac{235}{56}
,\\
f_{4,2}&=&
-\frac{4 x^5}{7}-\frac{25 x^4}{7}-\frac{17 x^3}{14}+\frac{171 x^2}{14}+\frac{153 x}{14}-\frac{25}{14}
,\\
f_{5,2}&=&
\frac{4 x^4}{7}+\frac{13 x^3}{7}-\frac{93 x^2}{14}+\frac{34 x}{7}-\frac{25}{14}
,\\
f_{6,2}&=&
\frac{1}{7} (-24) \left(2 \text{Li}_2\left(\frac{1-x}{2}\right)+\ln \left(\frac{x+1}{4}\right) \ln (x-1)\right)
.
\end{eqnarray}
For $l=3$, we have
\begin{eqnarray}
	f_{1,3}&=&
	\frac{20 x^7}{3}+40 x^6-\frac{20 x^5}{7}-\frac{375 x^4}{4}-\frac{220 x^3}{7}+\frac{185 x^2}{4}+20 x
	,\\
	f_{2,3}&=&
	\frac{25}{56} x (127 \log (x-1)+\log (x+1))
	,\\
	f_{3,3}&=&
	-\frac{20 x^7}{3}+\frac{8195 x^6}{168}+\frac{5225 x^5}{72}-\frac{35845 x^4}{252}-\frac{9977 x^3}{84}+\frac{7305 x^2}{56}+\frac{8941 x}{168}-32
	,\\
	f_{4,3}&=&
	-\frac{10 x^6}{3}-\frac{70 x^5}{3}-\frac{925 x^4}{42}+\frac{3425 x^3}{42}+\frac{195 x^2}{2}-\frac{845 x}{42}-\frac{635}{21}
	,\\
	f_{5,3}&=&
	\frac{10 x^5}{3}+\frac{40 x^4}{3}-\frac{1315 x^3}{42}+\frac{20 x^2}{7}+\frac{135 x}{14}+\frac{5}{21}
	,\\
	f_{6,3}&=&
	\frac{1}{7} (-200) x \left(2 \text{Li}_2\left(\frac{1-x}{2}\right)+\ln \left(\frac{x+1}{4}\right) \ln (x-1)\right)
	.
\end{eqnarray}
For $l=4$, we have
\begin{eqnarray}
	f_{1,4}&=&
	\frac{595 x^8}{22}+175 x^7-\frac{277315 x^6}{4928}-\frac{14305 x^5}{32}-\frac{202325 x^4}{4928}+\frac{14405 x^3}{48}\nonumber\\&&+\frac{342515 x^2}{4928}-\frac{1225 x}{32}-\frac{615}{64}
	,\\
	f_{2,4}&=&
	\frac{25 \left(7 x^2-1\right) (7613 \log (x-1)+67 \log (x+1))}{4928}
	,\\
	f_{3,4}&=&
	-\frac{595 x^8}{22}+\frac{1159715 x^7}{6336}+\frac{14823115 x^6}{44352}-\frac{8024815 x^5}{14784}-\frac{3123931 x^4}{4928}\nonumber\\&&+\frac{7841513 x^3}{14784}+\frac{169207 x^2}{448}-\frac{7430627 x}{44352}-\frac{2160637}{44352}
	,\\
	f_{4,4}&=&
	-\frac{595 x^7}{44}-\frac{4445 x^6}{44}-\frac{38235 x^5}{308}+\frac{113875 x^4}{308}+\frac{154025 x^3}{308}-\frac{42865 x^2}{308}\nonumber\\&&-\frac{6715 x}{28}-\frac{3735}{308}
	,\\
	f_{5,4}&=&
	\frac{595 x^6}{44}+\frac{665 x^5}{11}-\frac{34285 x^4}{308}-\frac{4540 x^3}{77}+\frac{30455 x^2}{308}+\frac{535 x}{77}-\frac{3735}{308}
	,\\
	f_{6,4}&=&
	\frac{1}{77} (-1500) \left(7 x^2-1\right) \left(2 \text{Li}_2\left(\frac{1-x}{2}\right)+\ln \left(\frac{x+1}{4}\right) \ln (x-1)\right)
	.
\end{eqnarray}
Here, $\text{Li}_n(z)$ is the polylogarithm function.

To obtain $C_{1,l},~C_{2,l},~C_{3,l},$ and $C_{4,l}$, we follow the method illustrated in Hinderer \cite{Hinderer:2007mb}. The general results have the following pattern:
\begin{eqnarray}
	C_{1,l}&=&A_{1,l}+\lambda_l B_{1,l},\\
	C_{2,l}&=&A_{2,l}+\lambda_l B_{2,l},\\
	C_{3,l}&=&A_{3,l}+\lambda_l B_{3,l},\\
	C_{4,l}&=&A_{4,l}+\lambda_l B_{4,l}.
\end{eqnarray} 
Here, $\lambda_l$ is related to the Love number $k_l$.
The constants also have the following pattern: $A_{i,l}\propto \mathcal{E}_m (GM)^l$ and $B_{i,l}\propto \mathcal{E}_m (GM)^{-l-1}$ $(i=1,2,3,4)$, where $\mathcal{E}_m$ is related to the static external quadrupolar tidal field produced by external gravitational potential. Subjected to this gravitational potential, the star responds through its own quadrupole moment, which is recorded by $\lambda_l$.

To obtain $\lambda_l$ and get rid of $\mathcal{E}_m$, we define $y(R)=R F'(R)/F(R)$ and $c=GM/R$. By substituting
\begin{equation}
F(x)=(C_{1,l}+\varepsilon C_{3,l}) Q^2_l(x) 
+(C_{2,l}+\varepsilon C_{4,l}) P^2_l(x)
+\varepsilon C_{2,l} S^2_l(x)
+\varepsilon C_{1,l} T^2_l(x)
\end{equation}
into $y$, we obtain
\begin{equation}
\lambda_l=-\frac
{
(A_{1,l}+\varepsilon A_{3,l})Q^*(R)
+(A_{2,l}+\varepsilon A_{4,l})P^*(R)
+\varepsilon\left(
A_{1,l}T^*(R)+A_{2,l}S^*(R)
\right)
}
{
(B_{1,l}+\varepsilon B_{3,l})Q^*(R)
+(B_{2,l}+\varepsilon B_{4,l})P^*(R)
+\varepsilon\left(
B_{1,l}T^*(R)+B_{2,l}S^*(R)
\right)
},\label{eq:LoveNumber0}
\end{equation}
where
\begin{eqnarray}
Q^*(R)&=&yQ^2_l(c)+c [d{Q^2_l}(c)/dc],\\
P^*(R)&=&yP^2_l(c)+c [d{P^2_l}(c)/dc],\\
T^*(R)&=&yT^2_l(c)+c [d{T^2_l}(c)/dc],\\
S^*(R)&=&yS^2_l(c)+c [d{S^2_l}(c)/dc].
\end{eqnarray}
Then, we can obtain the Love number through
\begin{equation}
k_l={(2l-1)!!\over 2R^{2l+1}}\lambda_l.
\end{equation}

Notice that because $\lambda_l\propto a_{i,l}/b_{i,l}\propto (GM)^{2l+1}$, then $\lambda_l R^{-2l-1} \propto c^{2l+1}$. Thus, we can redefine $k_l$ with 
\begin{equation}
	k_l=-{(2l-1)!!\over 2}\frac
	{
		(a_{1,l}+\varepsilon a_{3,l})Q^*(R)
		+(a_{2,l}+\varepsilon a_{4,l})P^*(R)
		+\varepsilon\left(
		a_{1,l}T^*(R)+a_{2,l}S^*(R)
		\right)
	}
	{
		(b_{1,l}+\varepsilon b_{3,l})Q^*(R)
		+(b_{2,l}+\varepsilon b_{4,l})P^*(R)
		+\varepsilon\left(
		b_{1,l}T^*(R)+b_{2,l}S^*(R)
		\right)
	}, \label{eq:LoveNumber}
\end{equation}
where $a_{i,l}$ and $b_{i,l}$ are just functions of $c$. We use Eq. \eqref{eq:LoveNumber} because the form is more straightforward to write in code than Eq. \eqref{eq:LoveNumber0}. In explicit form, the constants are shown below:
\begin{eqnarray}
a_{1,2}=0 ,&& b_{1,2}=\frac{15}{8} {1\over c^3},\\
a_{2,2}=\frac{1}{3} c^2,&& b_{2,2}= 0,\\
a_{3,2}=\frac{113}{84} c^2,&& b_{3,2}=\frac{1787}{392} {1\over c^3},\\
a_{4,2}=\frac{13}{9} c^2,&& b_{4,2}=-\left(\frac{5 \pi ^2}{7}+\frac{3305}{448}+\frac{15 \ln ^2(2)}{7}\right) {1\over c^3},
\end{eqnarray}
\begin{eqnarray}
a_{1,3}= 0,&& b_{1,3}=\frac{35}{8} {1\over c^4},\\
a_{2,3}=\frac{1}{45} c^3,&& b_{2,3}= 0,\\
a_{3,3}=\frac{127}{756} c^3,&& b_{3,3}=\frac{24805}{1512} {1\over c^4},\\
a_{4,3}=\frac{158}{945} c^3,&& b_{4,3}=-\left(\frac{25 \pi ^2}{9}+\frac{13795}{576}+\frac{25 \ln ^2(2)}{3}\right) {1\over c^4},
\end{eqnarray}
\begin{eqnarray}
a_{1,4}=0 ,&& b_{1,4}=\frac{735}{64 } {1\over c^5},\\
a_{2,4}=\frac{1}{630} c^4,&& b_{2,4}=0 ,\\
a_{3,4}=\frac{7613}{465696} c^4,&& b_{3,4}=\frac{469685}{5808} {1\over c^5},\\
a_{4,4}=\frac{200077}{10866240} c^4,&& b_{4,4}=-\left(\frac{875 \pi ^2}{88}+\frac{14680085}{202752}+\frac{2625 \ln ^2(2)}{88}\right) {1\over c^5}.
\end{eqnarray}

\subsubsection{Non-vacuum case}

From Eq. \eqref{eq:LoveNumber}, we need $y(R), M=m(R),$ and $r=R$ as inputs to obtain the Love number. Thus, we need to run the calculation of $F(r)$ in the interior.

From Eq. \ref{eq:0}, we obtain the following metric relations:
\begin{eqnarray}
	d=r^2/(ab),~e^\nu=e^\beta a/b^3,~e^{\bar{\lambda}}=e^\alpha/(ab),~\\
	H_0=H,~H_2=-H a^2/b^2,~F_0=-F_2=F.
\end{eqnarray}
We also need to define the perturbed stress tensor as
\begin{equation}
	T^\mu_\nu=T^\mu_{0~~\nu}+\delta T^\mu_\nu,
\end{equation}
with $T^\mu_{0~~\nu}$ as the usual ideal fluid and 
\begin{equation}
\delta T^\mu_\nu=\text{diag}\left(-{d\epsilon\over dp} ,1,1,1\right)
\delta p~Y_{lm}(\theta ,\phi ).
\end{equation}

After evaluating Eq. \eqref{eq:0.1} order by order and using identities of the spherical harmonics, one can arrive at the following relations:
\begin{eqnarray}
H&=&{16\pi G\kappa \delta p\over
b^2-a^2},\\
K&=&{\bar{G}\over ab}+{4\pi G\kappa\over ab}\left({1\over b^2}-{1\over a^2}{d\epsilon\over dp}\right)\delta p,\\
F&=&4\pi G \kappa\left(
{4\over b^2-a^2}-{3\over b^2}-{1\over a^2}{d\epsilon\over dp}
\right)\delta p
\end{eqnarray} 
Then, by eliminating $\delta p$, we have the following equations:
\begin{eqnarray}
	H&=&{4\over a^2-b^2}\left(
	{4\over a^2-b^2}+{3\over b^2}+{1\over a^2}{d\epsilon\over dp}
	\right)^{-1}F,\\
	F_1&=&C e^{(\alpha-\beta)/2},\\
	H_1&=&F_1\left(
	{a\over b^3}-{l(l+1)\kappa\over 2r^2}
	\right).
\end{eqnarray}
Lastly, we have
\begin{eqnarray}
F''&+&e^\alpha\left(
{2\over r}+{r\over 2\kappa}\left(
-4+{a\over b^3}+{3\over ab}
\right)
\right) F' 
-\left(
{2\over r^2}-{e^\alpha\over\kappa}
\left(
2-{3a\over b^3}+{1\over ab}
\right)
\right) F\nonumber\\
&+&e^\alpha \left({(l-1)(l+2)\over r^2}-{2\over ab\kappa}\right)G+K\left({2e^\alpha\over \kappa}\right)=0.
\end{eqnarray}
After some algebraic manipulations, the equation becomes
\begin{equation}
	F''+f(r) F'
	+g(r)F=0,
\end{equation}
with
\begin{eqnarray}
f(r)&=&\frac{r e^{\alpha (r)} }{\kappa }\left(\frac{1}{ab}-1\right)+\frac{e^{\alpha (r)}}{r}+\frac{1}{r},\\
g(r)&=&\frac{2 e^{\alpha} }{\kappa}{a\over b^3}\left(2-\frac{4}{\left(a^2-b^2\right) }\left(\frac{4}{a^2-b^2}+\frac{1}{a^2}{d\epsilon\over dp}+\frac{3}{b^2}\right)^{-1}\right)\nonumber\\
&&-\left(\frac{l(l+1) e^{\alpha }}{r^2}+\frac{2 e^{\alpha }}{\kappa }+\beta'(r)^2\right).
\end{eqnarray}
Defining $y(r)=r F'(r)/F(r)$, we thus obtain the first-order equation:
\begin{equation}
y'(r)=-f(r) y(r)-r g(r)-\frac{y(r)^2}{r}+\frac{y(r)}{r}.
\label{eq:Fnonvacuum}
\end{equation}
The boundary condition is $y(0)=l$.
Usually, $k_2$ (using Eq. \eqref{eq:LoveNumber}) is only evaluated as the so-called dimensionless tidal deformability $\Lambda$:
\begin{equation}
\Lambda={2k_2\over 3c^5},
\end{equation}
because $|k_4|\ll |k_3|\ll |k_2|$.

The numerical procedure for the tidal calculation is as follows: First, we calculate all $p'(r)$, $m'(r)$, and $y'(r)$ (Eqs.~\eqref{eq:p'},~\eqref{eq:m'} and~\eqref{eq:Fnonvacuum}, respectively). We employ the Runge-Kutta 4th-order algorithm using a {\it FORTRAN77} code. The initial data at the center $r=r_c\to 0$ are $p(r_c)=p_c$, $m(r_c)=0$ and $y(r_c)=l$. We run the code up to $r=R$ where the pressure becomes zero $p(R)=0$. At this point, we obtain $R$, $m(R)=M$, and $y(R)$. The three numbers are then used to calculate $k_l$ using Eq.~\eqref{eq:LoveNumber0}.

\section{NUMERICAL RESULTS AND DISCUSSIONS}
\label{result}
\begin{figure}
	\centering
	\includegraphics[width=0.5\linewidth]{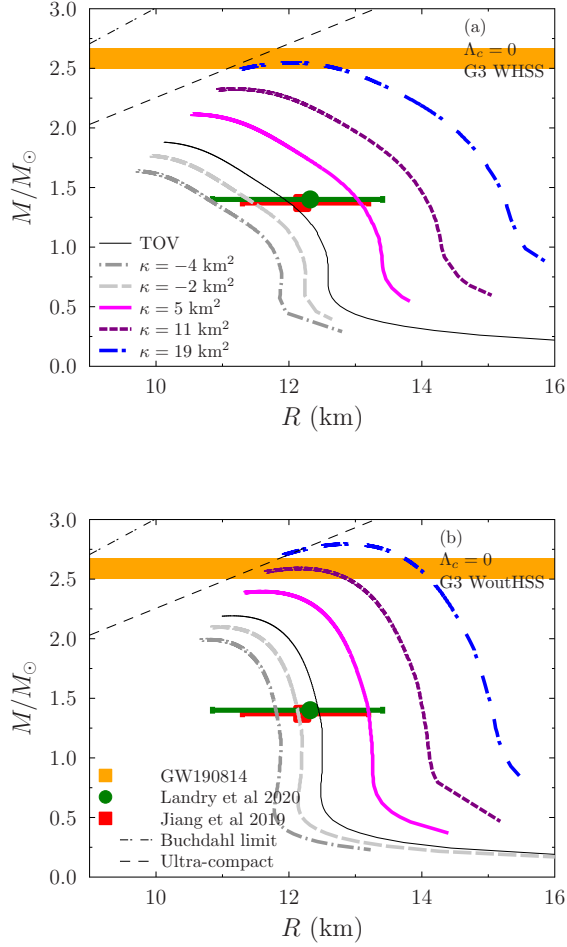}
	\caption{M-R relation results by varying $\kappa$ with $\Lambda_c=0$. In panel (a), we use G3 EoS with a hyperon and sound-of-speed constraint (WHSS), whereas in panel (b), we use no hyperon but still with the sound-of-speed constraint (WoutHSS). We can see that increasing $\kappa$ will increase $M$ and $R$. Similar to Ref.~\cite{Qauli:2016vza}, we found that $\kappa$ can have a negative value, and the solutions exist if $\kappa>-5$ km$^2$.}
	\label{fig:plotvarkappa}
\end{figure}

\begin{figure}
	\centering
	\includegraphics[width=0.5\linewidth]{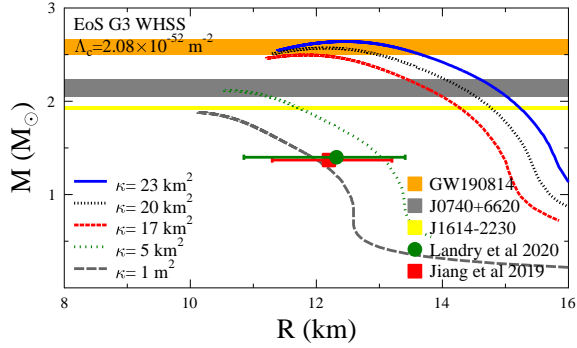}
	\caption{M-R relation from the realistic case $\Lambda_c=2.08\times 10^{-52}$ m$^{-2}$ is compared to the data from Landry \textcolor{black}{\it et al.} (2020)~\cite{Landry:2020vaw}, Jiang \textcolor{black}{\it et al.} (2019)~\cite{Jiang:2019rcw}, and pulsar-binary systems (PSRs) J1614-2230~\cite{Demorest:2010bx,Fonseca:2016tux,Arzoumanian:2017puf} and J0740+6620~\cite{Arzoumanian:2017puf,Cromartie:2019kug}. The GW190814 data were obtained from Ref.~\cite{Abbott:2020khf}.}
	\label{fig:M_R_3_2}
\end{figure}

\begin{figure}
	\centering
	\includegraphics[width=0.5\linewidth]{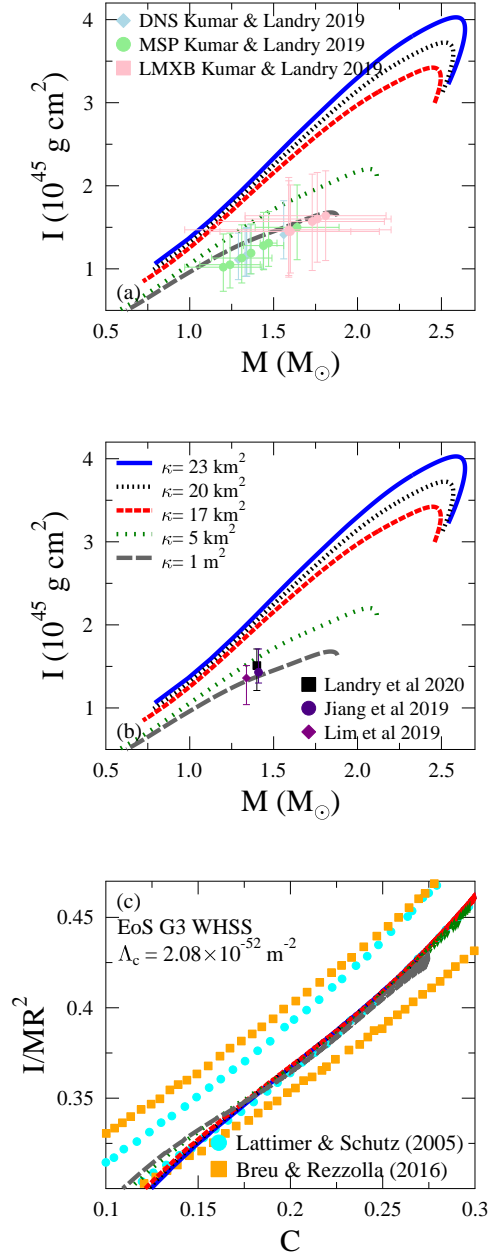}
	\caption{Moment of inertia $I$ from the realistic case $\Lambda_c=2.08\times 10^{-52}$ m$^{-2}$. In panel (a), we compare them with data from Kumar and Landry (2019)~\cite{Kumar:2019xgp}. The DNS, MSP, and LMXB correspond to \textcolor{black}{different astrophysical system}. In panel (b), we use Landry \textcolor{black}{\it et al.} (2020)~\cite{Landry:2020vaw}, Jiang \textcolor{black}{\it et al.} (2019)~\cite{Jiang:2019rcw}, and Lim \textcolor{black}{\it et al.} (2019)~\cite{Lim:2018xne}, whose error bars are quite narrow. In panel (c), we compare $I/MR^2$ with the upper and lower bounds from Lattimer and Schutz 2005~\cite{Lattimer:2004nj} and Breu and Rezolla 2016~\cite{Breu:2016ufb}.}
	\label{fig:MI_3}
      \end{figure}

 \begin{figure}
	\centering
	\includegraphics[width=0.5\linewidth]{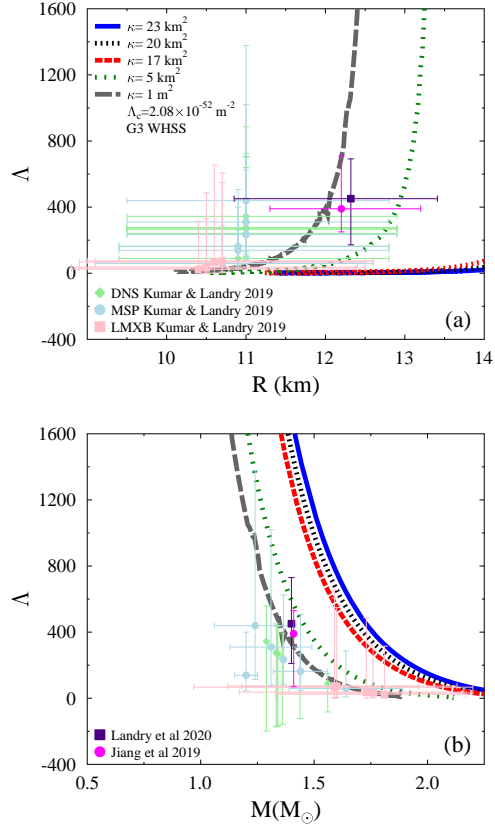}
	\caption{Dimensionless tidal deformability $\Lambda$ from G3 WHSS EoS with $\kappa$ varied for  $\Lambda_c=2.08\times 10^{-52}$ m$^{-2}$. In panel (a), we compare them with Kumar and Landry (2019)~\cite{Kumar:2019xgp}. In panel (b), we use Landry \textcolor{black}{\it et al.} (2020)~\cite{Landry:2020vaw} and Jiang \textcolor{black}{\it et al.} (2019)~\cite{Jiang:2019rcw}.}
	\label{fig:TD_3}
      \end{figure}

  \begin{figure}
  	\centering
  	\includegraphics[width=0.5\linewidth]{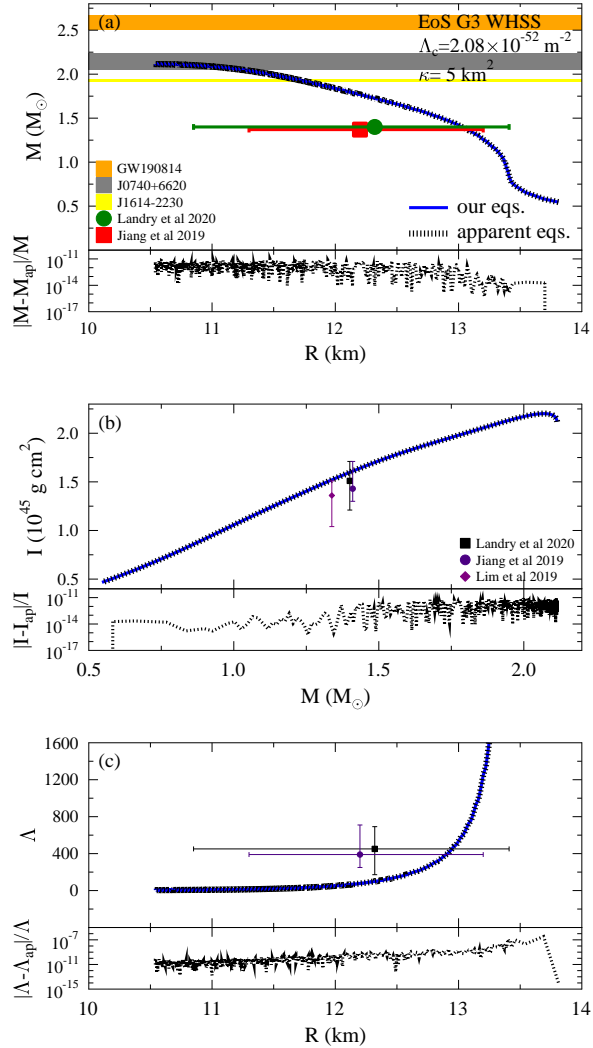}
  	\caption{\textcolor{black}{Here we compare the results from our calculations (our eqs.) versus the ones from employing the apparent EoS formulation (apparent eqs.)~\cite{Sham:2013cya}.}}
  	\label{fig:bandingap}
  \end{figure}

\begin{figure}
	\centering
	\includegraphics[width=0.5\linewidth]{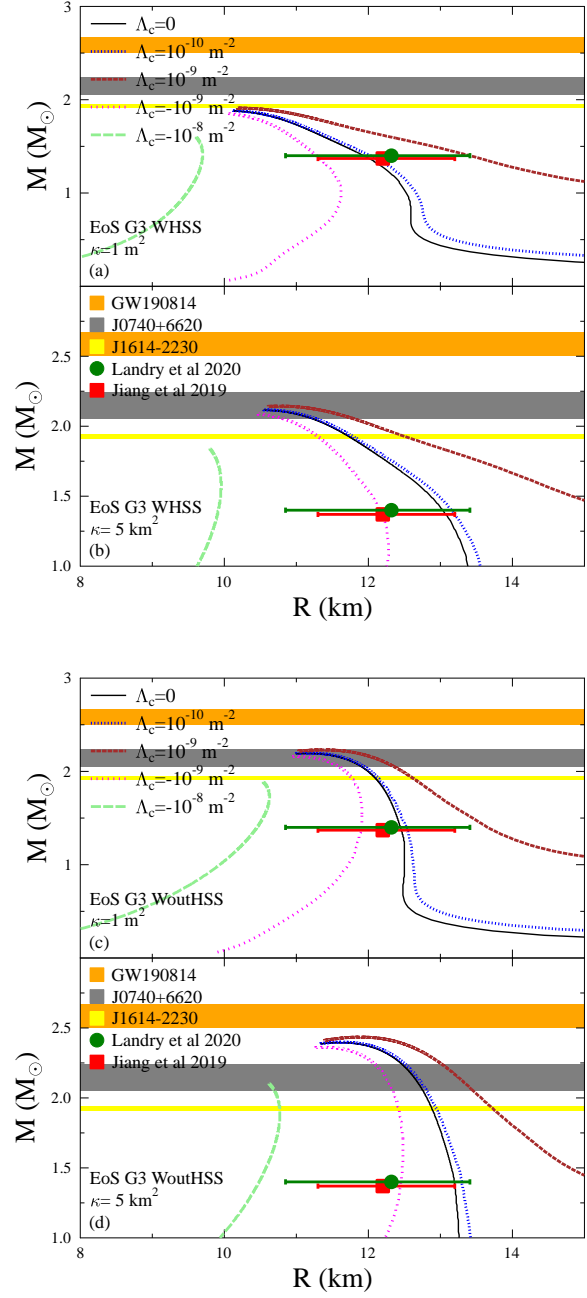}
	\caption{M-R relation results by varying $\Lambda_c$. In panels (a) and (b), we use small $\kappa$ ($\kappa=1$ m$^2$) and large $\kappa$ ($\kappa=5$ km$^2$), respectively, from the G3 WHSS EoS. Panels (c) and (d) contain the same thing except the EoS, which is WoutHSS.} 
	\label{fig:M_R} 
\end{figure}

\begin{figure}
	\centering
	\includegraphics[width=0.5\linewidth]{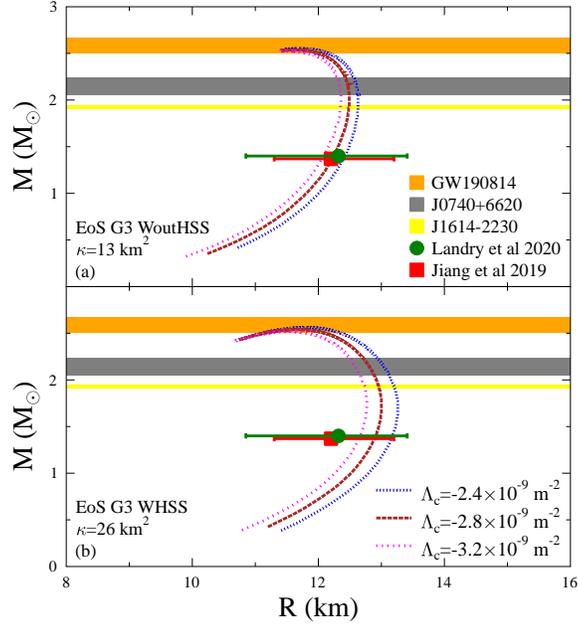}
	\caption{M-R curves from different EoSs. The curves that satisfy the data from Landry \textit{et al.} (2020) and Jiang \textit{et al.} (2019) are chosen, whose maximum mass also satisfies the GW190814 data. In panels (a) and (b), we use the EoS from G3 with and without hyperons, respectively, alongside the speed-of-sound constraint (WHSS and WoutHSS).}
	\label{fig:M_R_2}
\end{figure}

\begin{figure}
	\centering
	\includegraphics[width=0.5\linewidth]{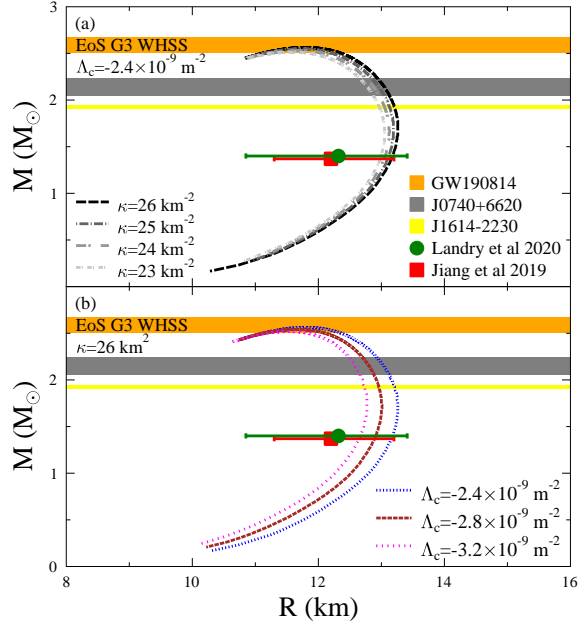}
	\caption{M-R curves from G3 WHSS with variations of $\kappa$ or $\Lambda_c$. These parameters' values are chosen to satisfy the data from Landry \textit{et al.} (2020) and Jiang \textit{et al.} (2019) while also maintaining their maximum mass at the range provided by the GW190814 data. In panels (a) and (b), we vary $\kappa$ and $\Lambda_c$, respectively.}
	\label{fig:M_R_3}
\end{figure}     

\begin{figure}
	\centering
	\includegraphics[width=0.5\linewidth]{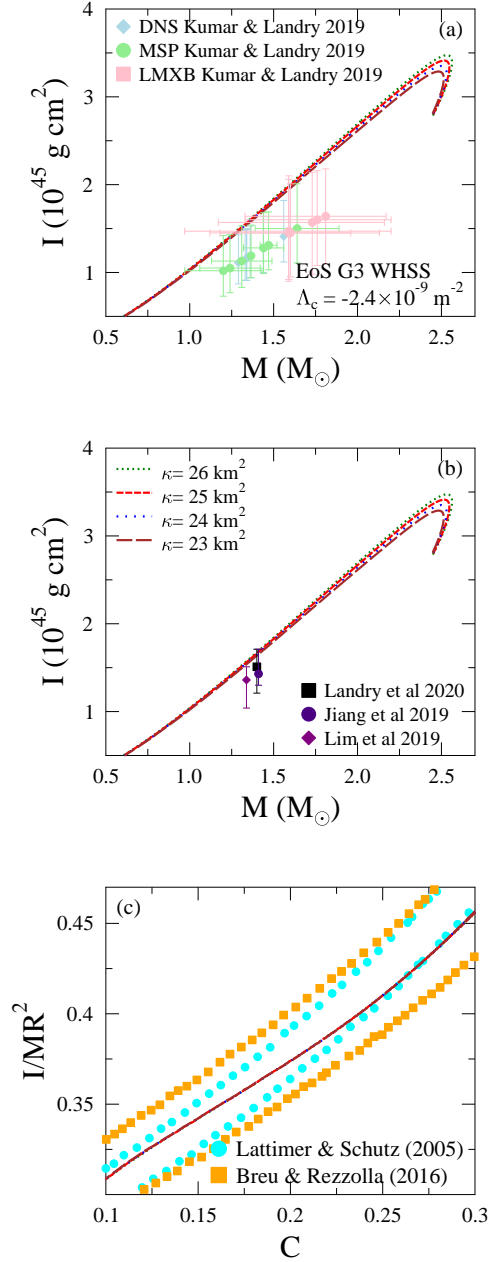}
	\caption{Moment of inertia within the EiBI gravity theory \textcolor{black}{obtained} from the G3 WHSS EoS and very large and negative $\Lambda_c$ by varying $\kappa$.}
	\label{fig:MI}
\end{figure} 
  \begin{figure}
	\centering
	\includegraphics[width=0.6\linewidth]{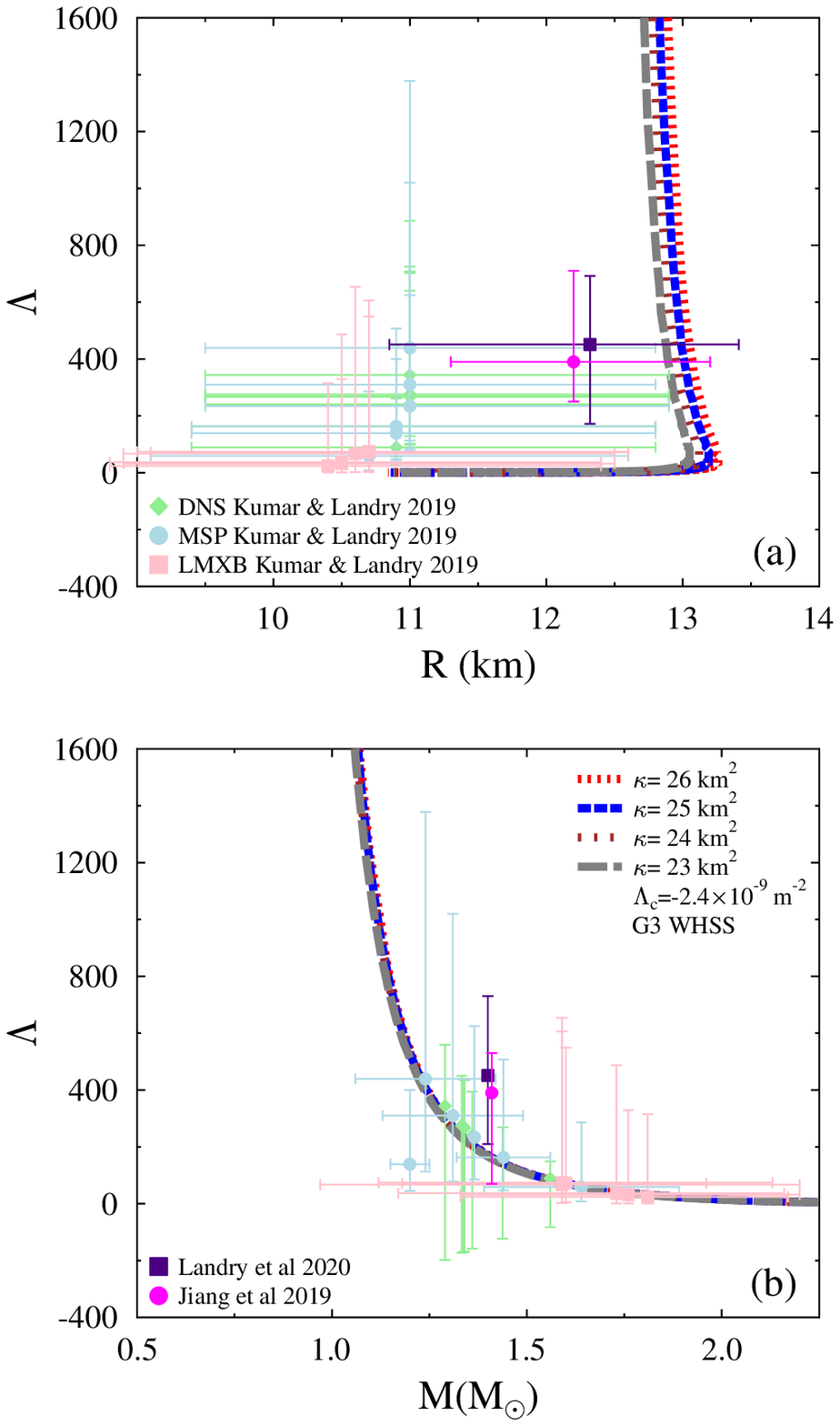}
	\caption{Tidal deformation with negative and very large values of $\Lambda_c$ and $\kappa$.}
	\label{fig:TD}       
\end{figure}

In this section, we show the numerical results. The EoS used here is the G3 parameter set. We include the hyperon contribution in the EoS with the speed of sound at high densities constrained by $v_s\leq c/\sqrt{3}$ (denoted by WHSS). We mainly use WHSS in this work, but we also compare the corresponding results with those with no-hyperon EoS (WoutHSS). We use the one hyperon contribution because we suspect that hyperon matter exists in heavy NSs, 
i.e., $M\gtrsim 2.0M_\odot$. We use the speed-of-sound constraint at high densities
because it stiffens the matter and thus increases the maximum mass. This effect can slightly reduce the impact of hyperon contributions. The hyperon contribution softens the EoS quite significantly, thus decreasing the maximum mass. Furthermore, the known \textcolor{black}{constraints from some analysis} results are shown as a comparison to restrict the range of $\kappa$ and $\Lambda_c$.  
The role of $\kappa$ for controlling the mass and radius of NS was studied previously in Ref.~\cite{Qauli:2016vza}. However, here, we revisit this matter using more refined EoS and use more recent constraints of NS properties. Then, we investigate the role of $\Lambda_c$.
	
First, we consider the case when $\Lambda_c$=0. The mass-radius relations are shown in Fig.~\ref{fig:plotvarkappa}, where the G3 WHSS EoS is presented in panel (a) and WoutHSS EoS in panel (b). In Fig.~\ref{fig:plotvarkappa}, we show that by increasing $\kappa$, the $M$ and $R$ of NS simultaneously increase. The impact of increasing $\kappa$ in increasing $M$ and $R$ is significant not only for the G3 WoutHSS EoS but also for the G3 WHSS EoS. The $M\sim 2.6M_\odot$ can be easily reached by the maximum mass predicted by both EoSs without crossing the Buchdahl limit. 
The results show that the G3 WHSS EoS needs a larger value of $\kappa$ to reach $M\sim 2.6M_\odot$ due to a relatively softer EoS. It is also evident that for EoSs with and without hyperons, the $2.1M_\odot$ maximum mass constraint and the radius canonical mass constraints from Refs.~\cite{Landry:2020vaw,Jiang:2019rcw} can be fulfilled simultaneously. However, when the maximum mass $M_{\rm max} \gtrsim 2.3M_\odot$, the radius of the canonical mass NS predicted by the EiBI theory is already larger than the other constraints~\cite{Landry:2020vaw,Jiang:2019rcw}.

We can estimate the upper bound of $\Lambda_c=2.08 \times 10^{-52}$ m$^{-2}$ from the observed cosmological constant in Refs.~\cite{Weinberg:1988cp,Carroll:2000fy,Padmanabhan:2002ji,Frieman:2008sn}
\begin{equation}
\rho_{\Lambda_c}=\frac{\Lambda_c}{8\pi G}\sim 10^{-8} \frac{\text{erg}}{\text{cm}^3}.
\end{equation}
This $\Lambda_c$ value is very small to provide an unphysical impact on the solar system. For example, when we use $|\Lambda_c|\geq 10^{-22}$ m$^{-2}$, $\lambda=1$, set $M=1 M_\odot$ and $r=1$ AU in Eq. \eqref{eq:3}, the cosmological constant term will dominate, and the Newtonian gravity will break down in the solar system. Second, we calculate the M-R relation, moment of inertia, and tidal deformability using this $\Lambda_c$ value. The results are almost indistinguishable compared to those obtained using $\Lambda_c=0$. We vary $\kappa$ and show the M-R relation, moment inertia, and tidal deformation results in Figs.~\ref{fig:M_R_3_2}-\ref{fig:TD_3}. Clearly, the data from NSs with canonical mass ($1.4 M_\odot$), such as moment of inertia and tidal deformation, are not in agreement with our results when $\kappa$ is much larger than $5$ km$^2$. 
\textcolor{black}{It is known that there is no direct measurement of both moment of inertia and tidal deformation. The data from Kumar and Landry (2019)~\cite{Kumar:2019xgp} is obtained by assuming some relations establised in GR, thus the data may be inappropriate to be used to test EiBI theory since the relations may be different in EiBI theory. On the other hand, the data from Landry \textcolor{black}{\it et al.} (2020)~\cite{Landry:2020vaw} and Jiang \textcolor{black}{\it et al.} (2019)~\cite{Jiang:2019rcw} come directly from gravitational wave observations thus is still valid in EiBI theory.} 
Moreover, when the cosmological constant is set to this value and $\kappa \approx$  $5$ km$^2$, the $2.1 M_\odot$ maximum mass constraint can be reached due to our choice of EoS. To this end, \textcolor{black}{restricted only to the EoS from G3 RMF parameter set,} we can conclude that to physically save the $\Lambda_c$ value, i.e., 0 $\le \Lambda_c \le$ 2.08 $\times 10^{-52}$ m$^{-2}$ and $\kappa \approx$  $5$ km$^2$, the NS properties predicted by the EiBI theory are compatible with the recent constraints from Refs.~\cite{Landry:2020vaw,Jiang:2019rcw,Arzoumanian:2017puf,Cromartie:2019kug,Kumar:2019xgp,Lim:2018xne}.

\textcolor{black}{We note that Sham \textcolor{black}{\it et al.} (2014)~\cite{Sham:2013cya} had investigated the moment inertia and tidal deformability of compact objects in EiBI. The authors in~\cite{Sham:2013cya}  employ the apparent EoS formulation in their calculation. This formulation makes the equations easier to derive because we can derive them just like the ones in GR but by replacing the energy density and pressure with apparent energy density and apparent pressure (Eqs.~\ref{eq:edenapp} and~\ref{eq:pressapp}). Here, we check the consistency of our results by comparing these results with the ones obtained by using the apparent EoS formulation. We show the comparison of both numerical results in Fig.~\ref{fig:bandingap}, where we use the case of G3 WHSS EoS, $\Lambda_c=2.08 \times 10^{-52}$ m$^{-2}$, and $\kappa = 5$ km$^2$. To make the difference more clearly, we also shown in the inset figure in lower part of each panel in Fig.~\ref{fig:bandingap}  the discrepancy of both formulations in $\Lambda$, $I$, and $M$ plots. Note that lower index $\rm ap$ in each corresponding quantity means the result obtained using apparent EOS formulation. It is evident from Fig. ~\ref{fig:bandingap} that both formulations are compatible. }

One tempting question: is it possible to reach the maximum mass of approximately $2.6 M_\odot$ while keeping the results still in agreement with the canonical mass observation data? After systematically studying all possible combinations of $\kappa$ and $\Lambda_c$, we have found that the case is only possible if we take the unphysical value of $\Lambda_c$, i.e., it should be negative and the $\Lambda_c$ absolute value should be much larger than $10^{-52}$ m$^{-2}$. The reasons are as follows: Increasing (decreasing) the value of $\Lambda_c$ affects the ``tail,'' corresponding to the MR curve on the lower right. If $\Lambda_c>0$, then the tail goes to the right. If $\Lambda_c<0$, then the tail goes to the left. For $\kappa=1 ~$m$^2$ and $\kappa=5 ~$km$^2$ cases and for both EoSs (WHSS and WoutHSS), the results of varying $\Lambda_c$ are shown in Fig.~\ref{fig:M_R}. Clearly, the impact of varying $\Lambda_c$ on the radius is greater than that of the $\kappa$ variation, except when near the maximum mass. A positive value of $\Lambda_c$ tends to increase the radius, whereas a negative value $\Lambda_c$ tends to decrease the radius. We compare the EoS G3 WHSS and G3 WoutHSS by comparing the plots in the upper and lower panels of Fig.~\ref{fig:M_R}. When we increase $\Lambda_c$, $R$ and $M$ increase and vice-versa. The plots show that for $\kappa=1 ~$m$^2$ and $\kappa=5 ~$km$^2$ cases, the range of $\Lambda_c$ of the G3 WHSS and G3 WoutHSS EoSs can be constrained with NS of approximately $M\sim 2.0M_\odot$ and canonical mass radius observation constraints \cite{Landry:2020vaw,Jiang:2019rcw,Demorest:2010bx,Fonseca:2016tux,Arzoumanian:2017puf,Cromartie:2019kug}. The range for $\kappa=1 ~$m$^2$ is quite wide, i.e., $-10^{-7}<\Lambda_c/($m$^{-2})<10^{-8}$, and this range is relatively wider than that of $\kappa=5 ~$km$^2$, i.e., $-10^{-9}\leq \Lambda_c/($m$^{-2})\leq 10^{-10}$. Thus, it is possible to have a relatively large maximum mass, but the radius is still retained small by increasing $\kappa$ value and decreasing $\Lambda_c$ value. However, for large $\kappa$ values, the range of the $\Lambda_c$ value becomes narrower. As a result, we can obtain the maximum mass of approximately $2.6 M_\odot$ and satisfy the radius constraint for $1.4M_\odot$ NS from the observations if we set $\kappa=26$ km$^2$ and $\kappa=13$ km$^2$ for the G3 WHSS EoS and G3 WoutHSS EoS, respectively, with an unavoidably large and negative $\Lambda_c$ but with a narrow range, i.e., $\Lambda_c=-(2.4-3.2)\times 10^{-9}$ m$^{-2}$. Their M-R curves are shown in Fig.~\ref{fig:M_R_2}. Of course, the combination of $\kappa$ and $\Lambda_c$ values can be chosen quite arbitrarily, but canonical mass radii and maximum mass constraints cannot be satisfied simultaneously when $\Lambda_c \ge$ 0. Note that the WHSS EoS yields a more significant radius shifting by varying $\Lambda_c$ than that of the WoutHSS EoS. The reason is that $\Lambda_c$ is usually not by itself in the equations but rather in the form of $\lambda=\kappa\Lambda_c+1$. In Fig.~\ref{fig:M_R_3}, we show the sensitivity of $\kappa$ and $\Lambda_c$ variations around the narrow region where the M-R curves satisfy the radius constraint for $1.4M_\odot$ NSs from the observations~\cite{Landry:2020vaw,Jiang:2019rcw} and $M\sim 2.6M_\odot$ from GW190814~\cite{Abbott:2020khf}, respectively. Evidently, from the lower panel of Fig.~\ref{fig:M_R_3}, the radius is quite sensitive to the $\Lambda_c$ variation. For completeness, we show the impact of the $\kappa$ variation on the \textcolor{black}{moment of inertia} and tidal deformation in Figs.~\ref{fig:MI} and~\ref{fig:TD}. For this case, the \textcolor{black}{moment of inertia} and tidal deformation are not too sensitive with the $\kappa$ variation, and the results are quite compatible with the NS results of the tidal deformability observations from Refs.~\cite{Kumar:2019xgp,Landry:2020vaw,Jiang:2019rcw,Abbott2017,Abbott2018,Abbott:2020khf}. 
 
To this end, we must consider very carefully the later results. The requirement that the maximum mass should be approximately $2.6 M_\odot$ and the radius constraint from the canonical NS $R_{1.4 M_{\odot}}$ should be 11 km $\lesssim R_{1.4 M_{\odot}}\lesssim 13$ km can only be satisfied by the EiBI gravity theory if the absolute value of the cosmological constant is unphysically large and the sign is negative. However, our universe has a positive and tiny cosmological constant~\cite{Weinberg:1988cp,Carroll:2000fy,Padmanabhan:2002ji,Frieman:2008sn}. Therefore, \textcolor{black}{restricted to the EoS from G3 RMF parameter set,} we conclude that the secondary object with $2.6 M_\odot$ observed in the GW190814 event~\cite{Abbott:2020khf} is not likely a static NS or a slow-rotating NS within the EiBI theory.
\section{CONCLUSIONS}
\label{concl}	

In conclusion, motivated by the assumption that the secondary compact object with $2.6 M_\odot$ observed in the GW190814 event could be an NS, we have systematically investigated the role of parameters $\kappa$ and $\Lambda_c$ of the EiBI gravity theory on the NS mass-radius relation, moment of inertia, and tidal deformability in the slow-rotating limit. The EoS of the core of an NS is calculated using the RMF model with the G3 parameter set~\cite{Kumara:2017bti}, where the SU(3) prescription and hyperon potential depths~\cite{Tolos:2017} are used to determine the hyperon coupling constants. For the inner and outer crusts, we use the crust EoS obtained from Miyatsu {\it et al.}~\cite{MYN2013}. We also ensured that the speed of sound in the matter does not exceed $c$/$\sqrt{3}$ at high densities. The G3 parameter set predictions with the experimentally and observationally nuclear matter and NS-related properties, including the nuclear matter EoS at intermediate densities, are shown. We have found that the NS mass $M$ significantly depends on the value of $\kappa$. For a positive $\kappa$ value, the NS maximum mass tends to increase when the $\kappa$ value increases, whereas for a negative $\kappa$ value, the NS maximum mass tends to decrease when the absolute $\kappa$ value increases. Furthermore, the NS radius $R$ depends significantly on the value of $\Lambda_c$. For a positive $\Lambda_c$ value, the NS radius tends to increase when the $\Lambda_c$ value increases, whereas for a negative $\Lambda_c$ value, the NS radius tends to decrease when the absolute $\Lambda_c$ value increases. We have also found that for EoS G3 with hyperon+speed of the sound treatment at high densities (G3 WHSS), for $\kappa \approx 5 ~$km$^2$ and $\Lambda_c$ value $\lesssim$ upper-bound $\Lambda_c$ value, the mass-radius relation satisfies the NS $M\sim 2.0M_\odot$ and canonical mass-radius observation constraints, respectively \cite{Landry:2020vaw,Jiang:2019rcw,Demorest:2010bx,Fonseca:2016tux,Arzoumanian:2017puf,Cromartie:2019kug}. If we use EoS with hyperons being excluded (G3 WoutHSS), the constraints can be satisfied even with a smaller $\kappa$ value. Furthermore, G3 WHSS and G3 WoutHSS EoSs can satisfy the maximum mass requirement of approximately $M\sim 2.6M_\odot$ and recent observation analysis results~\cite{Kumar:2019xgp,Landry:2020vaw,Jiang:2019rcw,Abbott2017,Abbott2018,Abbott:2020khf}, respectively. However, for the latter case, the $\kappa$ value is relatively large and the $\Lambda_c$ value is unphysically large and negative. In conclusion, if our universe is gravitationally governed by the EiBI gravity\textcolor{black}{, the NS matter in its core is described by the EoS from G3 RMF parameter set,} and the accepted value of a physical cosmological constant value is very small and positive, then the secondary object with $2.6 M_\odot$ observed in the GW190814 event~\cite{Abbott:2020khf} is not likely a static NS or a slow-rotating NS. We do not, however, rule out the possibility that such object is \textcolor{black}{either described by other EoS than G3 parameter set,} a fast-rotating NS, or in other modified gravity framework. They indeed deserve more investigation.

\begin{acknowledgments}
	
This work is funded by Kemenristek/BRIN's Penelitian Disertasi Doktor (PDD) 2021 grant No. NKB-314/UN2.RST/HKP.05.00/2021.

\end{acknowledgments}

\end{document}